\documentclass{emulateapj} 
\usepackage{amsmath}
\usepackage{verbatim}
\usepackage{color}
\usepackage{graphicx}

\newcommand{\be}{\begin{eqnarray}}
\newcommand{\ee}{\end{eqnarray}}
\newcommand{\bi}{\begin{itemize}}
\newcommand{\ei}{\end{itemize}}

\begin{document}

\title{Non-Thermal Electron Acceleration in Low Mach Number Collisionless Shocks. \\I. Particle Energy Spectra and Acceleration Mechanism}
\author{
Xinyi Guo$^1$, Lorenzo Sironi$^{1,2}$ and Ramesh Narayan$^1$
}
\affil{$^1$Harvard-Smithsonian Center for Astrophysics, 60 Garden St., Cambridge, MA 02138, USA \\$^2$NASA Einstein Postdoctoral Fellow
}

\begin{abstract}
Electron acceleration to non-thermal energies in low Mach number ($M_s\lesssim 5$)
shocks is revealed by radio and X-ray observations of galaxy clusters and solar flares, 
but the electron acceleration mechanism remains poorly understood. 
Diffusive shock acceleration, also known as first-order Fermi acceleration, cannot be directly invoked to explain the acceleration of electrons. Rather, an additional mechanism is required to pre-accelerate the electrons from thermal to supra-thermal energies, so they can then participate in the Fermi process. 
In this work, we use two- and three-dimensional particle-in-cell plasma simulations to study 
electron acceleration in low Mach number shocks. We focus on the 
particle energy spectra and the acceleration mechanism in a reference run with $M_s = 3$ and a quasi-perpendicular pre-shock magnetic field. 
We find that about $15\%$ of the electrons can be efficiently accelerated, forming a non-thermal power-law tail 
in the energy spectrum with a slope of $p\simeq2.4$. 
Initially, thermal electrons are energized at the shock front via shock drift acceleration. The accelerated electrons are then reflected back upstream, where their interaction with the incoming flow generates magnetic waves. In turn, the waves scatter the electrons propagating upstream back toward the shock, for further energization via shock drift acceleration. In summary, the self-generated waves allow for repeated cycles of shock drift acceleration, similarly to a sustained Fermi-like process.  
This mechanism offers a natural solution to 
the conflict between the bright radio synchrotron emission observed from the outskirts 
of galaxy clusters and the low electron acceleration efficiency usually expected in low Mach number shocks.
\end{abstract}

\section{Introduction}
Collisionless shocks occur in a wide variety of astrophysical settings: examples include Earth's bow shock, and the solar wind termination shock, supernova remnant (SNR) shocks in the interstellar medium, and structure formation shocks in the intracluster medium (ICM).

Particle acceleration is often associated with collisionless shocks. 
For instance, it is widely believed that Galactic cosmic rays with energies up to $\sim10^{15}\rm \,eV$ are ions accelerated by SNR shocks \citep[e.g.][]{Gaisser1990}. 
The most successful mechanism for explaining ion acceleration is diffusive shock acceleration \citep[DSA;][]{Blandford1978,Bell1978,Drury1983,Blandford1987}, also known as first-order Fermi acceleration (Fermi acceleration for short, hereafter). In DSA, charged particles cross the shock back and forth as they scatter off plasma/magneto-hydrodynamic (MHD) waves existing ahead and behind the shock (in the upstream and downstream regions, respectively). Since the turbulence is convected roughly at the local flow speed, waves on the two sides of the shock  effectively move towards each other due to the velocity jump at the shock. Hence the charged particles gain energy at each shock crossing. 

While the DSA mechanism has been successful in explaining ion acceleration in various settings \citep{Blandford1987}, 
e.g. in the Earth's bow shock \citep[see][for review]{Burgess2012} and SNR shocks \citep[see][for review]{Reynolds2008}.
However, DSA cannot be straightforwardly invoked for
the acceleration of electrons. To participate in the DSA process, thermal electrons need to cross the shock front multiple times. However, due to their small mass, electron gyro radii are very small compared to the shock thickness, which is controlled by the ion gyro radius. Thus, without undergoing some pre-acceleration, thermal electrons are expected to be tied closely to magnetic field lines and to be convected downstream without undergoing any significant DSA. This is known as the electron injection problem. 

To understand electron acceleration in shocks, fully kinetic numerical simulations are essential to self-consistently capture the non-linear loop that links the accelerated particles -- that generate turbulent magnetic fields --  to the turbulence itself -- which in turn governs the particle acceleration. In recent years, particle-in-cell (PIC) methods \citep[e.g.][]{Birdsall1991} 
have been used to simulate these kinetic processes.

So far, most of the work has focused on high Mach number shocks, where the Mach number $M_s$ is defined as the ratio of the shock speed to the sound speed of the ambient medium.
Many authors have found that, in high Mach number shock, the shock surfing acceleration (SSA) mechanism can inject high-energy electrons into DSA \citep{Dieckmann2000, McClements2001, Hoshino2002,Schmitz2002,Amano2007,Matsumoto2012}. 
In the SSA, large-amplitude electrostatic waves are excited at the leading edge of the shock transition region by the Buneman instability, as a result of the interaction between the reflected ions and the incoming electrons. These electrostatic waves trap the incoming electrons in their electrostatic potential. As a consequence, the trapped electrons can  be effectively accelerated by the shock motional electric field (i.e., the electric field resulting from the motion of the magnetized upstream region toward the shock). 
However, most of the previous work was based on simulations with a modest ion-to-electron mass ratio. \cite{Riquelme2011} argued that the importance of SSA decreases as the ion-to-electron mass ratio increases toward the realistic value. In this limit, rather than SSA, they found that it is the growth of oblique whistler waves near the front of quasi-perpendicular shocks that can pre-accelerate the electrons for long-term DSA.

Electron acceleration in low Mach number ($M_s\lesssim 5$) shocks has been poorly understood so far. The injection mechanism is expected to be different because the Buneman instability, essential for trapping the electrons near the shock for the SSA process, cannot be triggered at low Mach numbers\citep{Matsumoto2012}. 

On the other hand, low Mach number shocks are of great astrophysical interest. \cite{Lin2003} observed electron acceleration above solar flare tops and footpoints using X-ray data from \textit{Yohkoh} and \textit{RHESSI}. In galaxy clusters, low Mach number shocks have been identified in X-ray images \citep[e.g.][]{Markevitch2002,Russell2010,Akamatsu2012} and through observations of radio synchrotron emission by relativistic electrons accelerated at the shock \citep[e.g.][]{Wilson1970, Fujita2001, Govoni2004, vanWeeren2010,Linder2014}.
In addition, low Mach number shocks have been hypothesized to be present ahead of the G2 cloud \citep{Naryan2012,Sadowski2013} and of the S2 star \citep{Giannios2013} when they interact with the hot accretion flow at the Galactic Center \citep{Yuan2014}.

A few recent studies have explored electron acceleration in low Mach number shocks. 
Using one-dimensional (1D) PIC simulations, \cite{Matsukiyo2011} found efficient shock drift acceleration (SDA, e.g. \cite{Wu1984}, \cite{Krauss1989}, \cite{Ball2001}, \cite{Mann2006}) in low Mach number shocks. 
In the SDA process, particles gain energy from the shock motional electric field while drifting along the shock surface due to the gradient of the magnetic field at the shock front. 
However, the 1D nature of the simulations of \cite{Matsukiyo2011} prohibits a self-consistent study of self-generated waves in the upstream, since the wave-vector is confined to be perpendicular to the shock plane. 
Their results certainly suggest that SDA could be a potential injection mechanism, but they have no direct evidence of electron Fermi acceleration. 
Similarly, \cite{Park2012} and \cite{Park2013} studied perpendicular and quasi-perpendicular low Mach number shocks and found efficient SDA. However they did not see any evidence for sustained Fermi acceleration. 

In this paper, we study electron acceleration in low Mach number shocks using fully kinetic 2D and 3D PIC simulations. We focus on results from a reference run 
where the upstream magnetic field is quasi-perpendicular to the shock normal. Yet, in a forthcoming paper \citep{Guoinprep} we show that the results presented here can be 
generalized to a wide range of obliquity angles. 
We find a self-consistent mechanism for electron Fermi acceleration in which electrons are injected by pre-heating via SDA. These pre-accelerated electrons self-generate magnetic waves in the upstream region, and the waves in turn facilitate Fermi acceleration. 
The organization of this paper is as follows. 
In Section \ref{sec:setup}, the simulation setup with the various physical and numerical parameters
is described. In Section \ref{sec:shockstructure}, the shock structure of the reference run is discussed.
In Section \ref{sec:acc}, we present particle energy spectra and study in detail the electron acceleration mechanism of the reference run. We conclude with a discussion in Section \ref{sec:summary}.
In a forthcoming paper \citep{Guoinprep} we will study in detail the nature of the upstream waves and explore the parameter dependence
of the electron energy spectrum and acceleration mechanism.

\section{Simulation Setup}
\label{sec:setup}
We perform numerical simulations using the 3D electromagnetic PIC code TRISTAN-MP \citep{Spitkovsky05}, which is a parallel version of the publicly available code TRISTAN (Buneman 1993, p.67) that was optimized for studying collisionless shocks. 

\begin{figure}[tbp]
\begin{center}
\includegraphics[width=0.5\textwidth]{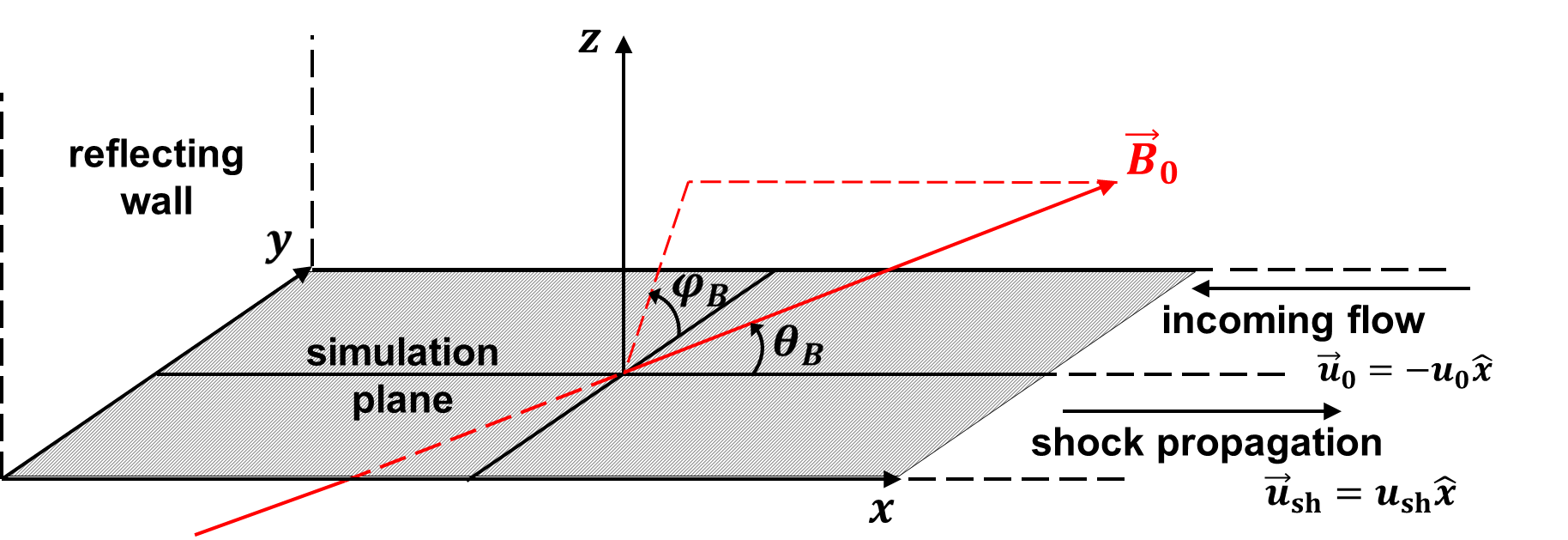}
\caption{Simulation setup. }
\label{fig:simplane}
\end{center}
\end{figure}

The computational setup and numerical scheme are described in detail in \cite{Spitkovsky2008,SS09,SS11,SSA13}. In brief, the shock is set up by reflecting an upstream electron-ion plasma, which follows a Maxwell-J\"uttner distribution with the electron temperature $T_e$ equal to the ion temperature $T_i$, and bulk velocity $\vec{u}_0 = -u_0\hat{x}$, off a conducting wall at the leftmost boundary $(x=0)$ of the computational box (Figure  \ref{fig:simplane}). 
The interplay between the reflected stream and incoming plasma causes a shock to form, which propagates along $+\hat{x}$ at the speed $u_{\rm sh}$. In the simulation frame, the downstream plasma is at rest.

The relation between the upstream bulk flow velocity and the plasma temperature is parametrized by the simulation-frame Mach number 
\begin{equation}
M \equiv \frac{u_0}{c_s} = \frac{u_0}{\sqrt{2\Gamma k_B T_i/m_i}}~,
\end{equation} where $c_s$ is the sound speed in the upstream, $k_B$ is the Boltzmann constant, and $\Gamma=5/3$ is the adiabatic index of the plasma, and $m_i$ 
is the mass of the ion.
The incoming plasma carries a uniform magnetic field $\vec{B}_0$, whose strength is parameterized by the magnetization
parameter 
\begin{equation}
\sigma\equiv \frac{B_0^2/4\pi}{ \left(\gamma_0-1\right)n m_ic^2 }~,
\end{equation}
where $\gamma_0 \equiv \left(1-u_0^2/c^2\right)^{-1/2}$
and $n=n_i = n_e$ is the number density of the incoming plasma.
The magnetic field orientation with respect to the shock normal (along $+\hat{x}$) is parameterized by the polar angle $\theta_{B}$ and azimuthal angle $\varphi_{B}$ (Figure \ref{fig:simplane}). 
The incoming plasma is initialized with zero electric field in its rest frame. Due to its bulk motion in the simulation frame, the upstream plasma carries a motional electric field $\vec{E}_0 = -(\vec{u}_0/c)\times\vec{B}_0$.

In the literature, the Mach number $M_s$ is often defined as the ratio between the upstream flow velocity and the upstream sound speed in the shock rest frame (rather than in the downstream frame, as in Equation 1). In the limit of 
weakly magnetized shocks, the Mach number $M_s$ is related to our simulation-frame Mach number $M$ through the implicit relation
\begin{equation}\label{eq:MsM}
M_s = M\frac{u_{\rm sh}^{\rm up}}{u_0}=M\left(1+\frac{1}{r\left(M_s\right)-1}\right),
\end{equation}
where $u_{\rm sh}^{\rm up}$ is the shock velocity in the upstream rest frame, equal to the upstream flow velocity in the shock rest frame, and 
\begin{equation}\label{eq:r}
r\left(M_s\right)= \frac{\Gamma+1}{\Gamma-1+2/M_s^2}
\end{equation}
is the Rankine-Hugoniot relation for the density jump from upstream to downstream.

For comparison with earlier work, where the magnetization 
is sometimes parametrized by the Alfv\'enic Mach number  $M_{\rm A}\equiv u_0/v_A$, where $v_A \equiv B_0/\sqrt{4\pi n m_i}$ is the Alfv\'en velocity, we remark that the relation between the magnetization and the Alfv\'enic Mach number is simply $M_A = \sqrt{2/\sigma }$. Alternatively, one could parameterize  the magnetic field strength by the plasma beta $\beta_{p} \equiv 8\pi nk_B\left(T_e + T_i\right)/B_0^2$, which is given by $\beta_p = 4/\left( \sigma\Gamma M^2\right)$ under the assumption of $T_e = T_i$.
We stress that in our simulations the upstream particles are
initialized with the physically-grounded Maxwell-J\"uttner distribution, instead of the so-called ``$\kappa$-distribution'' that was employed by, e.g., \cite{Park2013}.
The latter distribution artificially 
boosts the high energy component of the particle spectrum, thus artificially enhancing the acceleration efficiency in
most astrophysical settings (with the possible exception of shocks in solar flares, where the $\kappa$-distribution might be a realistic choice).

We perform simulations in both 2D and 3D computational domains. In our 2D simulations, we use a rectangular simulation box in the $xy$ plane, with periodic boundary conditions in the $y$ direction (Figure  \ref{fig:simplane}). In 3D, we employ periodic boundary conditions both in $y$ and in $z$. 
For both 2D and 3D domains, all three components of particle velocities and electromagnetic fields are tracked. As a result, the appropriate adiabatic index $\Gamma$ takes the value of $5/3$, as expected from a plasma with velocity in all three components. 
We find that most of the shock physics is well captured by 2D simulations, when the field is lying in the simulation plane, i.e. $\varphi_{B} = 0^\circ$. Therefore, to follow the shock evolution for longer times with fixed computational resources, we mainly utilize 2D runs, but we explicitly show in Appendix \ref{sec:2d3d} that our 2D results with an in-plane field configuration are in good agreement with full 3D simulations, while 2D results with an out-plane field configuration ($\varphi_B = 90^\circ$) do not agree with the 3D physics.

For accuracy and stability, PIC codes have to resolve the plasma oscillation frequency of the electrons 
\begin{equation}
\omega_{pe} = \sqrt{4\pi e^2n/m_e}~,
\end{equation} 
and the corresponding plasma skin depth $c/\omega_{pe}$, where $e$ and $m_e$ are the electron charge and mass. On the other hand, the shock structure is controlled by the ion Larmor radius 
\begin{equation}
r_{{\rm L},i}=\sqrt{\frac{2}{\sigma}}\sqrt{\frac{m_i}{m_e}}\ \frac{c}{\omega_{pe}}\gg \frac{c}{\omega_{pe}}~,
\end{equation} and the evolution of the shock occurs on a time scale given by the ion Larmor gyration period $\Omega_{ci}^{-1}=r_{{\rm L},i}u_0^{-1}\gg \omega_{pe}^{-1}$.
The need to resolve the electron scales, and at the same time to capture the shock evolution for many $\Omega_{ci}^{-1}$, is an enormous computational challenge, for the realistic  mass ratio $m_i/m_e=1836$. Therefore, we decide to employ a reduced mass ratio $m_i/m_e = 100$ for most of our runs. In Appendix \ref{sec:massratio}, we discuss in detail the dependence of our results on the mass ratio. We find that the results are in perfect agreement between $m_i/m_e =100$ and $m_i/m_e = 400$. Thus our results can be generalized to the realistic mass ratio of $m_i/m_e=1836$, with the scalings that we discuss in Appendix \ref{sec:massratio}.

\begin{figure*}[tbp]
\begin{center}
\includegraphics[height=0.6\textheight]{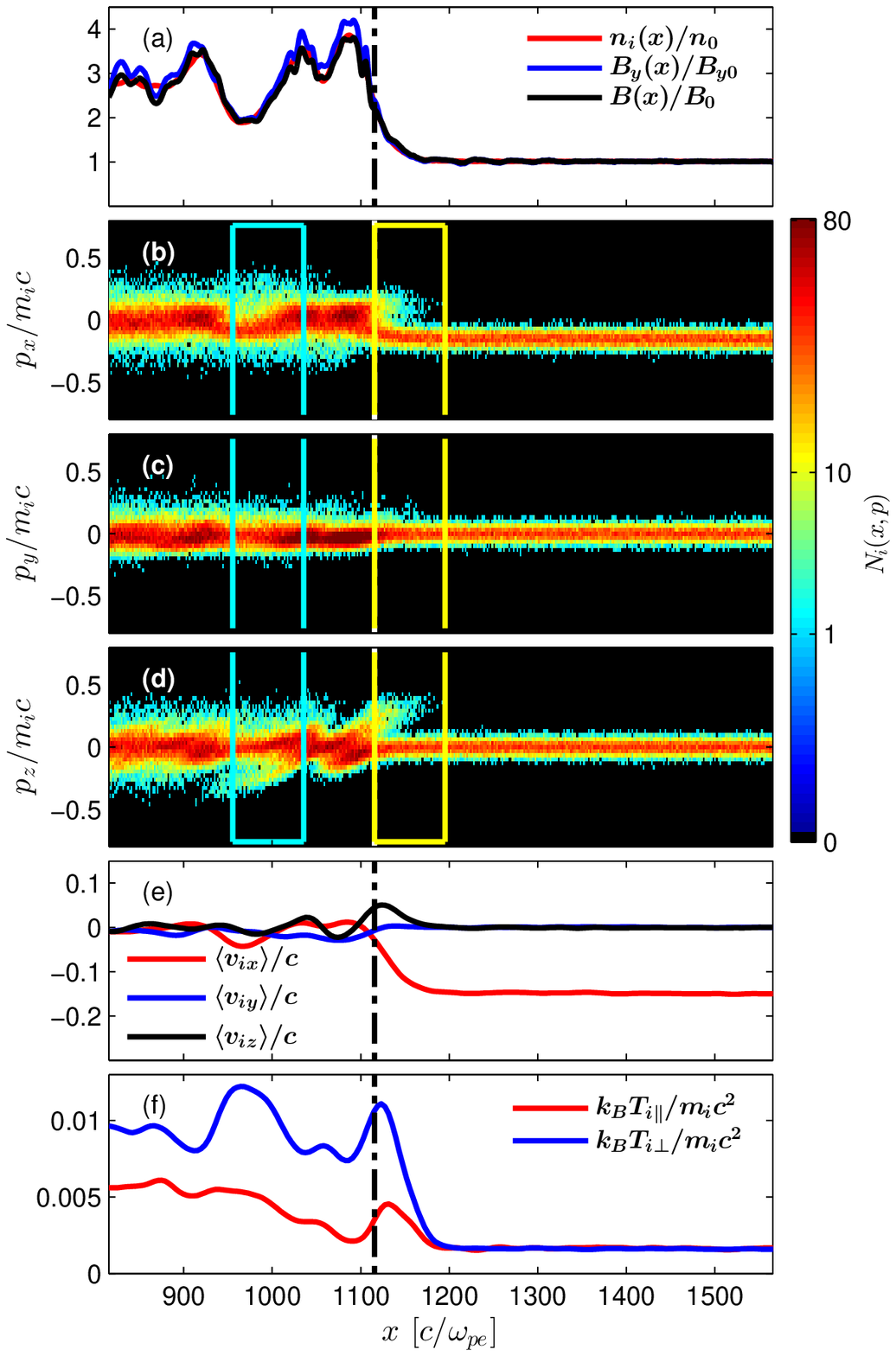}\includegraphics[height=0.6\textheight]{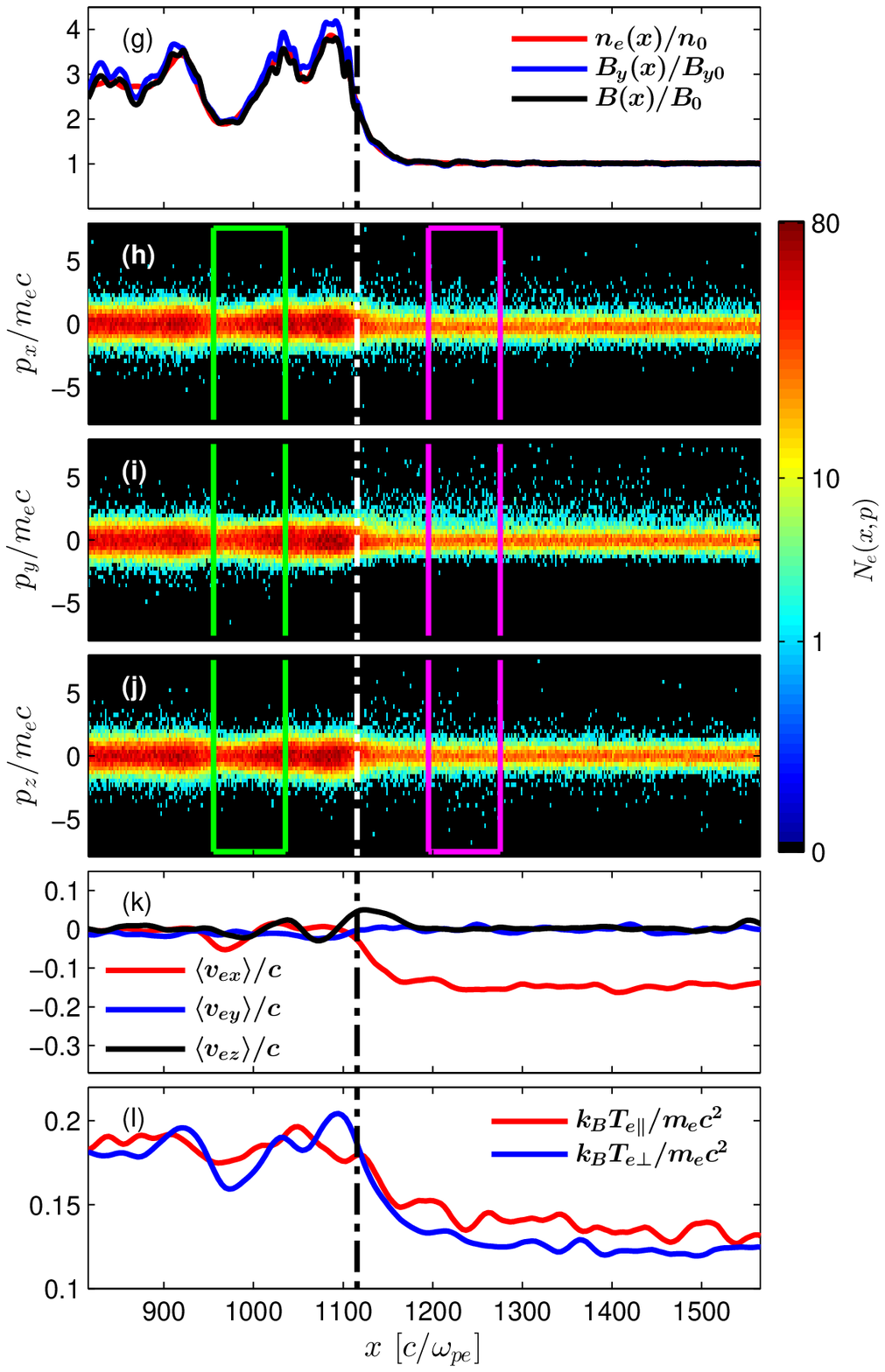}
\end{center}
\caption{ Shock structure of our reference run at time $\omega_{pe}t = 14625 \ (\Omega_{ci}t=26.9)$.
The shock is at $\simeq1115\ c/\omega_{pe}$, indicated by the vertical dot-dashed lines, 
and moves towards the right. The downstream is to the left of the shock, and the upstream to the right.
The left column shows quantities related to the ions and the right column quantities related to the electrons.
From top to bottom, we present the ratios $n/n_0$, $B_y/B_{y0}$, $B/B_0$, the momentum phase space plots
$p_x-x$, $p_y-x$, $p_z-x$, the $y$-averaged velocity profiles $\langle v_x\rangle,\langle v_y\rangle,\langle v_z\rangle$,
and the temperatures parallel ($T_{\parallel}$) and perpendicular ($T_{\perp}$) to the magnetic field.  The yellow box encloses $0-80\ c/\omega_{pe}\ (0-1\ r_{\rm L,i})$ ahead of the shock. The cyan box encloses $80-160\ c/\omega_{pe}\ (1-2\ r_{\rm L,i})$ behind the shock. The time evolution of the ion energy spectra in these two regions is shown in Figure \ref{fig:specevol}(a) and(b). The green and the magenta boxes enclose regions at $80-160\ c/\omega_{pe}$ behind the shock and $80-160\ c/\omega_{pe}$ ahead of the shock, respectively. The time evolution of the electron energy spectra in these two regions is shown in Figure \ref{fig:specevol}(c) and (d).
}
\label{fig:egshock}
\end{figure*}
To further optimize our use of computational resources, 
the incoming particles are initialized at a ``moving injector", which recedes from the wall in the $+\hat{x}$ direction at the speed of light. When the injector approaches the right boundary of the computational domain, we expand the box in the $+\hat{x}$ direction. 
This way both memory and computing time are saved, while following at all times the evolution of the upstream regions that are causally connected with the shock. Further numerical optimization can be achieved by allowing the moving injector to periodically jump backward (i.e. in the $-\hat{x}$ direction), resetting the fields to its right (see \cite{SS09}).
Since we expect the acceleration to happen close to the shock, we choose to jump the injector in the $-\hat{x}$ direction such that to keep a distance of at least a few tens of ion Larmor radii ahead of the shock. This suffices to properly capture the acceleration physics. 
We have checked that, albeit at relatively early times, simulations with and without the jumping injector show consistent results. 

In the main body of this paper, we present the results from a reference run simulated on a 2D domain. 
The upstream plasma in this reference run is initialized with $T_i = T_e = 10^9K = 86{\ \rm keV}/k_{B}$ and $u_0 = 0.15\,c$, which results in a 
simulation-frame Mach number $M=2$. Using Equation \ref{eq:MsM}, this corresponds to $M_s = 3$. 
The strength of the magnetic field is set so that the magnetization is $\sigma = 0.03$ and the field lies in the simulation plane at an oblique angle with respect to the shock normal, such that
$\theta_B = 63^\circ$ and $\varphi_B = 0^\circ$. 
The parameters are chosen to resemble closely the simulation 
presented in \cite{Naryan2012}, which studied a low Mach number shock ($M_s = 2$) that might be formed 
during the passage of the G2 cloud through the accretion disk at the Galactic Center. 
In this work, our choice of the Mach number ($M_s = 3$) and of the magnetization ($\sigma = 0.03$) is relevant also for  
shocks in galaxy clusters, where $M_s \sim 1.5-5$, $B_0 \sim 1\,\mu G$, $n\sim 10^{-4}-10^{-5}{\rm cm}^{-3}$ and 
$u_0 \sim 1000\ {\rm km/s}$ \citep{Matsukiyo2011}.  
We remark that although the chosen value for the plasma temperature ($T_i = T_e = 10^9K = 86\,{\rm keV}/k_{ B}$) 
is fairly high in the context of the plasma in ICM, where $k_BT \sim 10\ {\rm keV}$
for rich clusters, 
the acceleration physics does not change 
much with temperature. We will explicitly show the dependence of our results on the flow temperature in a forthcoming paper \citep{Guoinprep}. 

We employ a spatial resolution of $10$ cells per electron skin depth $c/\omega_{pe}$, have $32$ particles per cell ($16$ per species) and use
a time resolution of $dt = 0.045\ \omega_{pe}^{-1}$. 
The transverse box size is fixed at $76\ c/\omega_{pe}$ (corresponding to $7.6\ c/\omega_{pi}$, or about one ion Larmor radius). We have performed convergence tests which show that we can properly resolve the acceleration physics with 5 cells per $c/\omega_{pe}$, and we have confirmed that simulations with a number of particles per cell up to $64$ and a transverse box size up to $256\ c/\omega_{pe}$ give essentially the same results. 

\section{Shock Structure}
\label{sec:shockstructure}

\begin{figure*}[tbp]
\begin{center}
\includegraphics[width=0.7\textwidth]{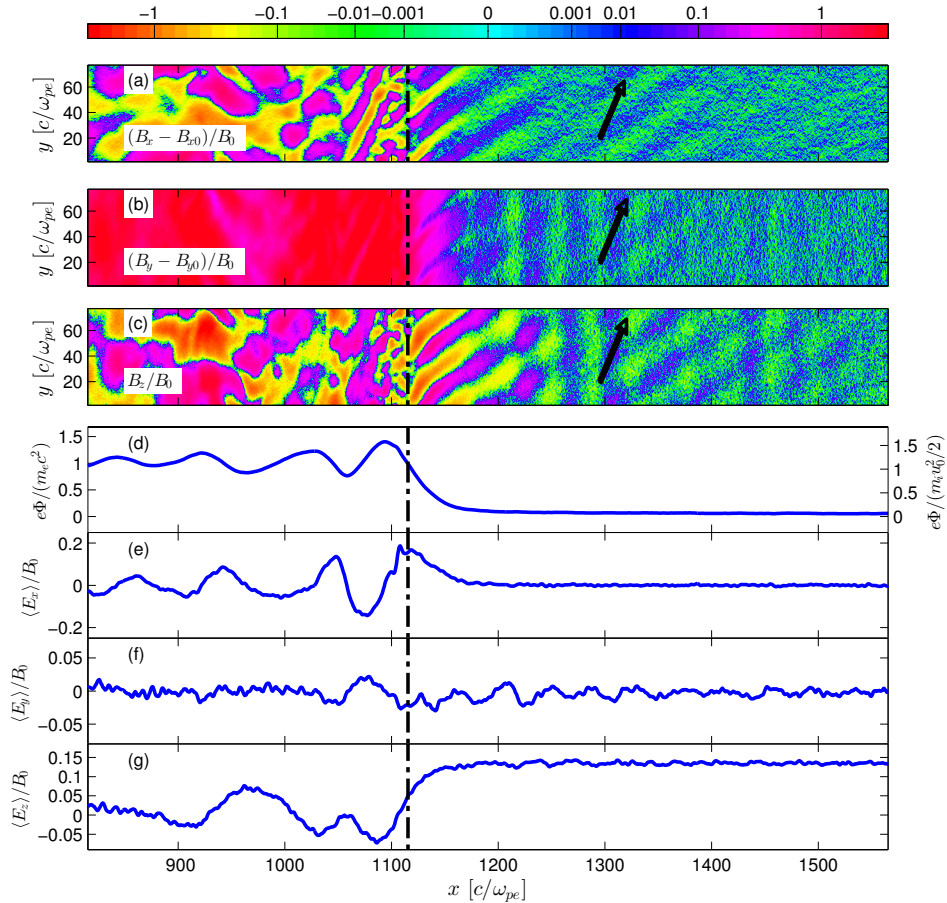}
\end{center}
\caption{Electromagnetic fields in the reference run at time $\omega_{pe}t = 14625 \ \left(\Omega_{ci}t =26.9\right)$. 
Panels (a)-(c) show 2D plots of the magnetic field components in units of $B_0$, after subtracting the background field $\vec{B}_0$ (i.e., we show $(B_x-B_{x,0})/B_0$, $(B_y-B_{y,0})/B_0$ and $B_z/B_0$, respectively).
 The black arrows indicate the orientation of the upstream background magnetic field $\vec{B}_0$. Note that there are waves in all three components 
of the magnetic field.
Panel (d) shows the $y$-averaged electric potential normalized by the electron rest mass energy, i.e. $e\Phi/\left(m_ec^2\right)$ (as indicated 
by the vertical axis on the left) or by the ion bulk kinetic energy, i.e. 
$e\Phi/\left(m_i u_0^2/2\right)$ (as indicated by the vertical axis on the right). 
Panels (e)-(g) show the $y$-averaged electric field components normalized by the upstream magnetic field strength, $\langle E_x\rangle/B_0$, $\langle E_y\rangle/B_0$, $\langle E_z\rangle/B_0$.
Since the upstream plasma is initialized with zero electric field in its rest frame, 
the expected motional electric field $\vec{E}_0 = -(\vec{u}_0/c)\times \vec{B}_0$ in the simulation frame 
in the upstream region is $\langle E_z\rangle/B_0=u_0/c\sin\theta_B  = 0.134$. 
}
\label{fig:shockflds}
\end{figure*}
In this section we present the shock structure of our reference run. 
Figure \ref{fig:egshock} shows quantities related to ions in the left column and those related to electrons 
in the right column. Each column shows 
the longitudinal profiles of number density $n$, transverse magnetic field $B_y$ and total magnetic field strength $B$, the momentum spaces $p_x-x$, $p_y-x$, $p_z-x$, the average velocity profiles $\langle v_x\rangle,\langle v_y\rangle,\langle v_z\rangle$, the temperature parallel to the magnetic field $T_{\parallel}$ and perpendicular to the field $T_{\perp}$.
Figure \ref{fig:shockflds}(a)-(c) shows 2D plots of the magnetic field components in units of $B_0$, after subtracting the background field $\vec{B}_0$ (i.e., we show $(B_x-B_{x,0})/B_0$, $(B_y-B_{y,0})/B_0$ and $B_z/B_0$, respectively).
These quantities provide a good characterization of the waves seen in the magnetic field.
Figure \ref{fig:shockflds}(d) shows the $y$-averaged value of the electric potential $\Phi(x) = -\int_{\infty}^{x} \langle E_x(x') \rangle dx'$ normalized by the electron rest mass energy [$e\Phi/(m_ec^2)$, vertical scale on the left] or by the ion bulk kinetic energy [$e\Phi/(m_i u_0^2/2)$, vertical scale on the right].\footnote{Strictly speaking, the quantity $\Phi$ we define is not the electric potential, but rather the electromotive force along the $x$ direction. Yet, for the sake of simplicity, in the following we refer to $\Phi$ as the electric potential.} Figure  \ref{fig:shockflds}(e)-(g) presents the $y$-averaged components of the electric field $\langle E_x\rangle$, $\langle E_y\rangle$, $\langle E_z\rangle$. 
All quantities are measured in the simulation frame at $\omega_{pe}t = 14625\ (\Omega_{ci}t=26.9)$. 
The shock front is at $x\simeq 1115\ c/\omega_{pe}$ (indicated by the vertical dot-dashed lines in all the panels) and moves towards the right with a velocity of $u_{\rm sh}\simeq 0.076\,c$.
This velocity yields $M_s = M\left( u_{\rm sh}^{\rm up}/u_0\right)
\sim M\left(u_{\rm sh}+u_0\right)/u_0 \simeq 3$, in agreement with Equation \eqref{eq:MsM}, that applies to a weakly magnetized medium.

In Figure \ref{fig:egshock}(a) and (g) we show the ion  and electron density normalized by the upstream value (red curves). The plasma density is compressed in the downstream,  as expected. The influence of the magnetic field on the shock jump conditions can be neglected when the upstream plasma beta is $\beta_p\gg 1$ \citep{Tidman1971}. 
For our reference run, we have $\beta_p = 20\gg1$, so the expected compression ratio, $r=n_2/n_1$, where $n_2$ and $n_1$ are the downstream and upstream plasma density respectively, can be safely estimated from the Rankine-Hugoniot conditions for an unmagnetized shock (Equation \eqref{eq:r}). 
With $M_s =3$ and $\Gamma=5/3$, the jump condition yields $r=3$. 

In the simulation, the compression ratio is $r\sim 4$ near the shock front and relaxes to $r\sim 3$ farther downstream, as expected. 
The compression of the magnetic field strength is closely related to the density compression, by the freezing of magnetic flux. Under the $\beta_p\gg 1$ condition, the compression of the magnetic field transverse to the shock normal ($B_y$ in this case) can be approximated by $r$. Indeed, in the simulation, we observe that the compression of $B_y$ (blue curve in Figure  \ref{fig:egshock}(a)) is almost identical to the density  compression (red curve in Figure  \ref{fig:egshock}(a)). Since the gradient of the total magnetic field strength drives the drift motion of electrons in the process of SDA (Section \ref{sec:SDA}), we also plot the profile of the total field $B$ (black curve). As the upstream magnetic field is quasi-perpendicular ($\theta_{B}=63^\circ$), the compression of total magnetic field is only slightly less than the compression of the transverse component.  

The overshoot of the compression ratio at the shock has been well studied, see, e.g., \cite{Leroy1981, WuWinske1984}. It is ascribed to a population of gyrating ions that stay close to the shock front. 
These ions are reflected by the ambipolar electric field (Figure \ref{fig:shockflds}(e)), induced by the different inertia of the ions and electrons entering the shock. The potential energy $e\Phi$ of the ions near the shock front is comparable
to the their bulk kinetic energy $m_i u_0^2/2$, as indicated by the vertical axis on the right side of Figure \ref{fig:shockflds}(d).\footnote{Both observations and numerical simulations of shocks show that $e\Phi\sim m_i u_0^2/2$ holds generally \citep{Amano2007}.}
It follows that a fraction of the ions having $m_i v_x^2/2<e\Phi$ will be reflected at the shock front, to form a stream of gyrating particles 
ahead of the shock. Upon reflection, they acquire energy as they drift towards the $+\hat{z}$ direction 
along the shock motional electric field (Figure  \ref{fig:shockflds}(g)). The resulting energy gain allows the gyrating ions to overcome the 
potential barrier and thus advect downstream at their second encounter with the shock. The gyromotion during 
their first encounter affects the magnetic field structure, and contributes to the formation of the magnetic overshoot seen in Figure  \ref{fig:egshock}(a). 

The existence of the reflected ions can be inferred from the ion phase space plots in Figure \ref{fig:egshock}(b)-(d)).
The reflected ions are located just ahead of the shock and they have large and positive values of $p_x$ and $p_z$. We note that the gyrostream formed by the reflected ions is confined within $0-1\ r_{\rm L,i}$ ahead of the shock (as delimited by the yellow box in Figure \ref{fig:egshock}). 
Beyond a few Larmor radii ahead of the shock,  
the ions follow the distribution at initialization, in terms of average velocities (Figure  \ref{fig:egshock}(e)) and temperature (Figure  \ref{fig:egshock}(f)).

In the downstream, the ions are heated anisotropically, with the momentum dispersion along the field (i.e., along the $y$ direction) being smaller.\footnote{In the absence of efficient pitch angle scattering, it is harder for the ions to isotropize along the direction of the mean field.} The resulting ion temperature anisotropy in the downstream is shown in Figure  \ref{fig:egshock}(f). 

Electrons, having the opposite charge with respect to ions, cannot be reflected by the potential barrier at the shock. Rather, the potential tends to pull them into the downstream region.
Yet, we observe a significant fraction of electrons reflecting back upstream, propagating far ahead of the shock. These are the electrons with very large momenta ($p_{x,y,z}\gtrsim 3\,m_ec$) ahead of the shock in the electron phase spce plots (Figure \ref{fig:egshock}(h)-(j)). 
The reflection happens, according to the theory of SDA (Section \ref{sec:SDA}), due to the jump of the magnetic field at the shock, which effectively acts as a magnetic mirror. These reflected electrons have preferentially large momenta parallel to the upstream magnetic field, as seen from the excess of electrons with large positive $p_y$ in the upstream (Figure \ref{fig:egshock}(i)). Since the returning electrons preferentially move along the upstream magnetic field, the
electron temperature parallel to the magnetic field $T_{e\parallel}$ is larger than the perpendicular component $T_{e\perp}$ in the upstream (Figure \ref{fig:egshock}(l)). 

The electron temperature anisotropy is closely related to the waves in 
the magnetic field shown ahead of the shock in Figure  \ref{fig:shockflds}(a)-(c). 
The waves in $B_z$ (Figure  \ref{fig:shockflds}(c)) show two oblique modes symmetric around the direction of the background magnetic field (indicated by the black arrow in Figure \ref{fig:shockflds}(a)-(c)). 
The waves in $B_x$ and $B_y$ tend to prefer one of the two oblique modes present in $B_z$ (Figure \ref{fig:shockflds}(a) and (b), respectively).
The oscillatory pattern in $\langle E_y\rangle/B_0$ (Figure  \ref{fig:shockflds}(f)) is
associated with the upstream magnetic waves in $B_z$, which are advected roughly at the upstream fluid velocity $\vec{u}_0= -0.15\,c\,\hat{x}$.
More precisely, by analyzing high time resolution data from the simulation, 
we measure that the average phase velocity of the waves is indeed $\vec{v}_{w}\simeq -0.15\,c\,\hat{x}$  in the simulation frame. The fact that the waves are moving toward the shock 
suggests that the particles triggering the waves must exist beyond the region where the waves are present. The waves clearly exist beyond a distance of a few hundred $c/\omega_{pe}$ ahead of the shock. So do the returning 
electrons that cause the upstream electron temperature anisotropy (Figure \ref{fig:egshock}(l)). On the other hand, there are no returning ions  beyond $80\ c/\omega_{pe}$ ahead of the shock,  as evident from the ion phase spaces, the average velocity profile and the temperature plots (Figure \ref{fig:egshock}(b)-(f)). This is a strong evidence that the upstream waves are driven by the returning electrons, not by the ions.

\section{Particle Energy Spectra and Acceleration Mechanism}
\label{sec:acc}
In this section, we first present the evolution of the energy spectra of ions and electrons in the upstream 
and downstream regions of the reference run. We show that the electron energy spectrum in the upstream develops
a clear non-thermal component and the high energy tail stretches in time to higher and higher energies, indicating efficient acceleration persisting over time. 
Then in subsection \ref{sec:subacc}, we describe the electron acceleration 
mechanism by identifying SDA as the injection mechanism 
and showing how it enables sustained Fermi acceleration. 

\subsection{Spectral Evolution}
\begin{figure*}[tbp]
\begin{center}
\includegraphics[width=0.9\textwidth]{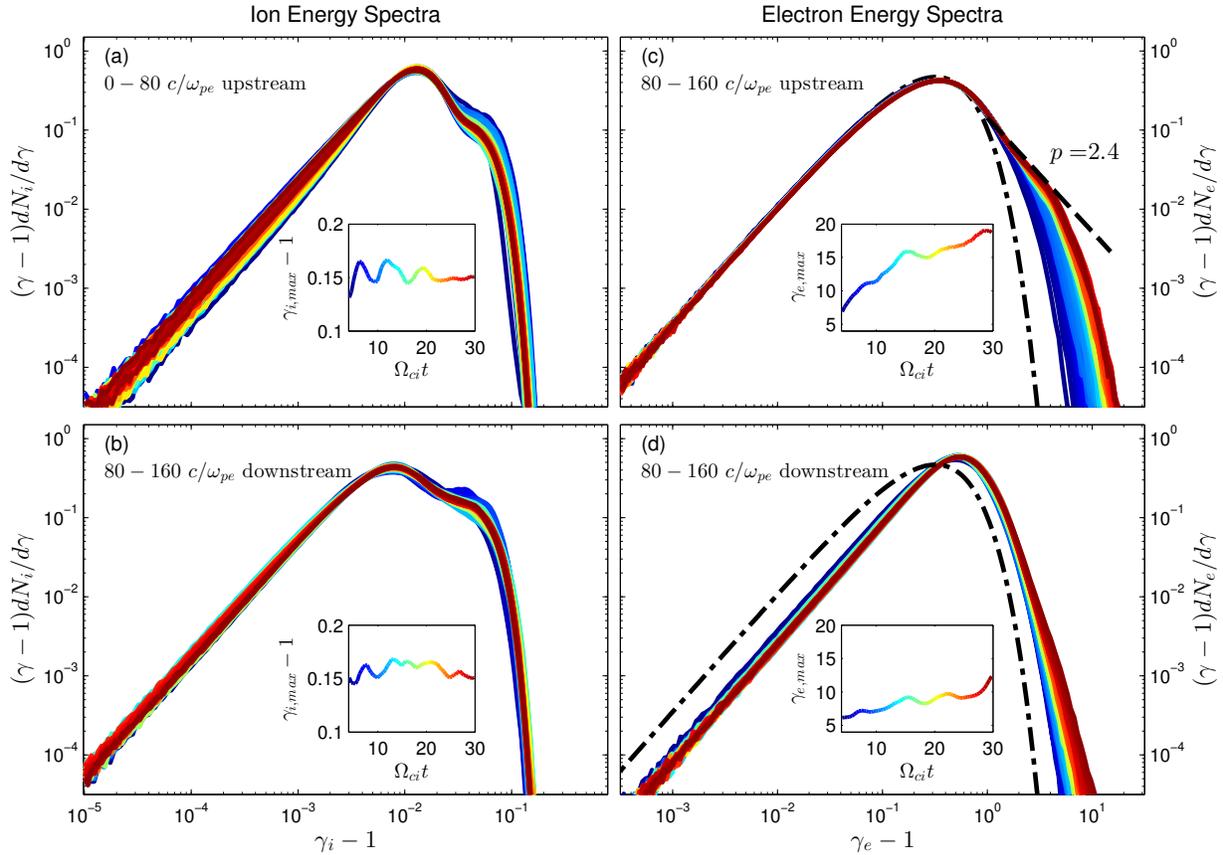}
\caption{ Time evolution of the ion (left) and electron (right) energy spectra from $\Omega_{ci}t=4.1$ up to $29.8$, measured in the upstream (top) or downstream (bottom) regions.
(a) ion energy spectra taken $0-80\ c/\omega_{pe}\ (0-1\ r_{\rm L,i})$ ahead of the shock (yellow box in Figure \ref{fig:egshock}). 
(b) ion energy spectra taken $80-160\ c/\omega_{pe}\ (1-2\ r_{\rm L,i})$ behind the shock (blue box in Figure \ref{fig:egshock}). 
(c) electron energy spectra taken $80-160\ c/\omega_{pe}$ ahead of the shock (magenta box in Figure \ref{fig:egshock}). 
(d) electron energy spectra taken $80-160\ c/\omega_{pe}$ behind the shock (green box in Figure \ref{fig:egshock}).
Color indicates time, from blue to red as the simulation evolves from early to late times.
The subplot of each panel traces the maximum particle energy over time for the particles in the same slab where the spectrum is computed. 
Dot-dashed lines in panels (c) and (d) represent the initial electron energy spectrum, namely, a  Maxwellian distribution with $T=10^9 K$ drifting at the bulk velocity $\vec{u}_0 = -0.15\,c\,\hat{x}$. 
The dashed line in panel (c) shows the best-fit power law of the 
non-thermal component in the late-time upstream electron energy spectrum.
Panels (a) and (b) show that the ion energy spectra barely evolve over time. In contrast, panels (c) and (d) show that the electrons continue to be accelerated to higher and higher energies over time.
}
\label{fig:specevol}
\end{center}
\end{figure*}

Figure \ref{fig:specevol} follows the time evolution of the ion and electron energy spectra at fixed distances from the shock in the upstream and downstream regions, as marked by the colored boxes in Figure \ref{fig:egshock}. 
At each instant in time, we define the maximum energy of ions or electrons
as the Lorentz factor $\gamma_{i,max}$ and $\gamma_{e,max}$ at which the particle number density drops below $10^{-4.5}$, the 
lowest level shown in our spectrum plots. 
(The value of $10^{-4.5}$ is arbitrary, but our results are not sensitive to this choice.)
We show in the subpanels of Figure  \ref{fig:specevol}
how $\gamma_{i,max}$ and $\gamma_{e, max}$ evolve over time. 

From the ion energy spectrum ahead of the shock in Figure \ref{fig:specevol}(a), we see that about $20\%$ of the incoming ions are reflected at the shock and form a high energy component in the upstream spectrum (at $\gamma_i-1\gtrsim 0.03$). The high-energy end of the ion spectrum shows little temporal evolution, as revealed also by the fact that the maximum Lorentz factor $\gamma_{i,max}$ is nearly constant over time. Similarly, the downstream ion energy spectrum in Figure \ref{fig:specevol}(b) shows little time evolution. In addition, we have confirmed that
the ion energy spectra taken further upstream $(x-x_{\rm shock}\gtrsim r_{\rm L,i}$, where $x_{\rm shock}$ is the shock position)  strictly follow the initial drifting Maxwellian distribution at all times, which demonstrates that no reflected ions can reach this region. This agrees with the ion phase space plots in Figure \ref{fig:egshock}(b)-(d). Based on the lack of evolution in the ion spectra, we conclude that the acceleration of ions will not proceed to higher energies. We point out that this conclusion is valid only in the 
quasi-perpendicular regime with $\theta_B\gtrsim 45^\circ$. For quasi-parallel field configurations $\left(\theta_B\lesssim 45^\circ\right)$, ions can be efficiently accelerated to higher and higher energies by DSA  \citep{Caprioli2014}. But an investigation of the physics of ion acceleration is beyond the scope of this paper.

In contrast to the ion energy spectrum, the upstream electron energy spectrum (Figure \ref{fig:specevol}(c)) clearly develops a non-thermal tail over time. The fractional number density of non-thermal electrons is roughly $15\%$ and the spectral index $p$, defined as $dN/d\gamma \propto (\gamma-1)^{-p}$, has a best-fit value $p=2.4$. 
The steady growth of the maximum electron energy $\gamma_{e, max}$ clearly shows that electron acceleration is efficient and persistent over time. 

The downstream electron spectrum (Figure  \ref{fig:specevol}(d)) initially follows a Maxwellian distribution, whose temperature is higher than in the upstream. After $\sim20\ \Omega_{ci}^{-1}$, 
a non-thermal component begins to develop and the maximum energy increases. This happens because some of the electrons that get accelerated in the upstream are eventually advected downstream after being reflected back toward the shock by the upstream waves (Section \ref{sec:Fermi}). However, the non-thermal tail in the downstream electron  spectrum evolves more slowly than in the upstream spectrum, and it is hard to disentangle from the Maxwellian distribution. Because of this, we will only study the upstream electron energy spectrum in the following sections.

\begin{figure*}[tbp]
\begin{center}
\includegraphics[width=0.7\textwidth]{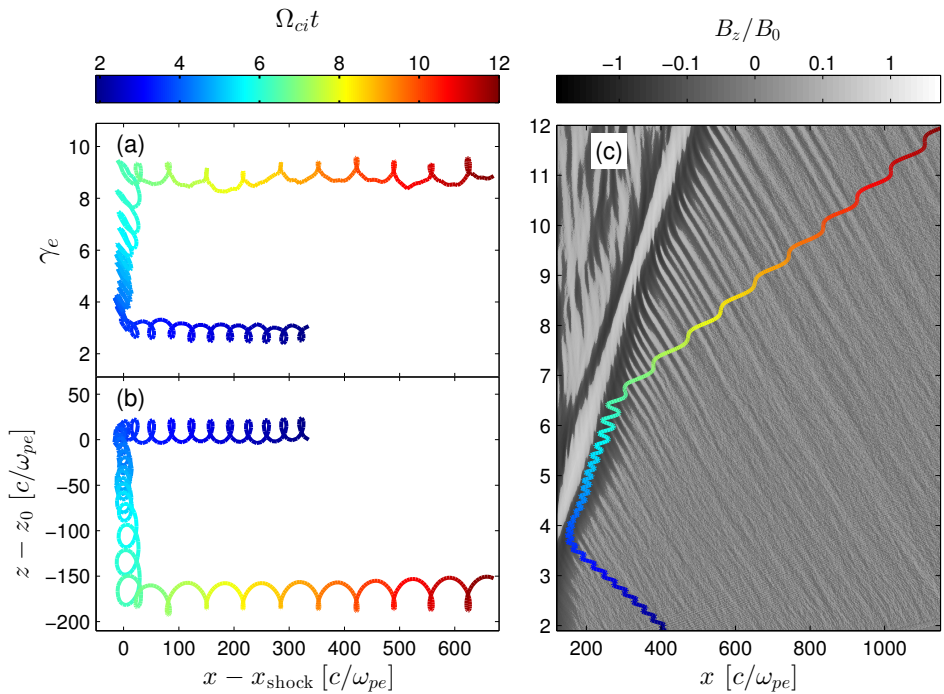}
\caption{
The evolution of a typical electron undergoing shock drift acceleration (SDA). 
Color indicates time, from blue to red as time evolves from early to late.
Panels (a) and (b) show the evolution of the electron Lorentz factor
 and $z$-coordinate as a function 
of its distance from the shock. 
Panel (c) shows the $x$-location of the electron on top of the spatio-temporal evolution of the $y$-averaged value of $B_z/B_0$. 
All quantities are measured in the simulation (downstream) frame. 
}
\label{fig:testprt_eg1}
\end{center}
\end{figure*}

\subsection{The Acceleration Process}\label{sec:subacc}
In general, the electron acceleration process operating in the reference run can be summarized as follows. 
First, a fraction of the incoming electrons, whose pitch angle and energy satisfy the reflection criteria of the shock-drift acceleration (SDA) process, which we shall describe in detail in Section \ref{sec:SDA}, 
will drift along the shock surface and gain energy from the motional electric field. 
After the energization at the shock front, they are reflected upstream with increased energy. Their momentum is preferentially oriented along the upstream magnetic field, causing an electron temperature anisotropy (with $T_{e\parallel}> T_{e\perp}$) in the upstream region. 
The induced anisotropy drives self-consistently the generation of upstream waves,
which can scatter the reflected electrons propagating upstream (after the SDA process) back toward the shock, for further energization through SDA.
We identify this self-sustained process as a form of Fermi acceleration. 

We first study the details of SDA in Section \ref{sec:SDA}.
We then illustrate the electron Fermi acceleration process by showcasing the evolution of
a typical non-thermal electron in Section \ref{sec:Fermi}. 

\subsubsection{Shock Drift Acceleration Theory}\label{sec:SDA}
The non-relativistic theory of SDA including non-zero cross-shock electric potential has been reviewed by \cite{Ball2001}. \cite{Mann2006} have developed the relativistic theory of SDA with vanishing cross-shock potential and \cite{Park2013} have studied the effect of non-zero cross-shock potential in the non-relativistic limit. In our reference run, the upstream electrons are hot and marginally relativistic, and the cross-shock potential is also significant compared to the electron rest mass energy. In this section, we summarize findings from previous work and generalize the fully relativistic SDA theory to properly treat arbitrary value of cross-shock potential. 

We begin by addressing the relation between the drift motion of electrons at the shock and their energy gain by the 
SDA mechanism. The physics can be intuitively understood in the downstream frame. 
The gradient of the magnetic field (along $-\hat{x}$) near the shock front (see Figure \ref{fig:egshock}(a)) causes an incoming electron to drift with 
$\vec{v}_{\nabla B}=(-p_{\perp}^2/2m_ee\gamma B^3)(\vec{B}\times \vec{\nabla} B)$
along the $-\hat{z}$ direction. Here, $p_\perp$ denotes the component of the particle momentum perpendicular to the magnetic field. The electron is then accelerated by the motional electric field $E_z$ (Figure \ref{fig:shockflds}(g))
in the downstream rest frame. The contribution to the electron energy gain by $E_z$ takes the form
\begin{equation}
\Delta \gamma_{\rm SDA} =\frac{-e}{m_ec^2}\int E_z\, dz~,
\end{equation}
where the integral is over the drift path. Assuming that the electric field near the shock is constant and is purely due to the upstream motional electric field, $E_z(x)\simeq E_0\equiv u_0/c B_0\sin\theta_B$, then the energy gain via SDA in the downstream frame can be estimated as
\begin{equation}\label{eq:deltagam}
\Delta\gamma_{\rm SDA} \simeq  \frac{-e}{m_e c^2} E_0\, \Delta z = -\sqrt{\frac{\sigma }{2}}\sqrt{\frac{m_i}{m_e}}\,\frac{u_0^2}{c^2}\sin\theta_B\,\frac{\Delta z}{c/\omega_{pe}}~.
\end{equation} 
The time scale of the SDA drift cycle is estimated by \cite{Krauss1989,Krauss1989b} to be $\sim \Omega_{ci}^{-1}$, which 
we have verified in our simulations.

Figure \ref{fig:testprt_eg1} shows the trajectory of an electron undergoing one cycle of SDA, at an early stage of the shock evolution. The SDA process operates from $\Omega_{ci}t \sim 3.5$ up to $\Omega_{ci}t\sim 6.5$, during which the 
particle stays within $\sim 50\ c/\omega_{pe}$ ahead of the shock. The electron drifts along $-\hat{z}$ for a distance $\Delta z \simeq -200\ c/\omega_{pe}$, and its energy gain is $\Delta \gamma \simeq 6$. 
Using the relevant parameters $\sigma = 0.03, u_0 = 0.15\,c, \theta_B = 63^\circ$, we find that the energy gain indeed comes almost exclusively from the drift motion along $-\hat{z}$, in agreement with Equation \eqref{eq:deltagam}. In other words, $\Delta \gamma\sim \Delta \gamma_{\rm SDA}$. After being reflected by the shock at $\Omega_{ci}t\sim 6.5$, the electron propagates back into the upstream.

In order to understand  the efficiency of SDA and the conditions under which an incoming 
electron can participate in SDA, it is more convenient to 
switch to two other frames: the de Hoffman-Teller (HT) frame \citep{HT1950} or the upstream rest frame.
The upstream rest frame is a convenient choice to analyze the properties of the incoming plasma. From the downstream rest frame, it is obtained by simply boosting with velocity ${u}_0$ along the shock normal.
The HT frame has the advantage that the motional electric 
field vanishes on both sides of the shock, since the flow velocity is parallel to the background magnetic field both ahead and behind the shock. The HT frame can be obtained, starting from the upstream rest frame, by boosting in the direction opposite to the upstream magnetic field with a velocity
\begin{align}\label{eq:betas}
u_t = u_{\rm sh}^{\rm up} \sec\theta_B &= \frac{u_{\rm sh}+u_0}{1+u_{\rm sh}u_0/c^2}\sec\theta_B~.  
\end{align}
Note that the HT frame can be defined only for shocks with $u_t\leq c$. Shocks with $u_t>c$ are characterized as superluminal, 
and the SDA physics explained below does not apply.
For the parameters of our reference run, magnetic field configurations with $\theta_B\gtrsim 77^\circ$ are superluminal.
Our choice of the obliquity angle $\theta_B = 63^\circ$ is thus well below the superluminality threshold.

\begin{figure*}[tbp]
\begin{center}
\includegraphics[width = 1.0\textwidth]{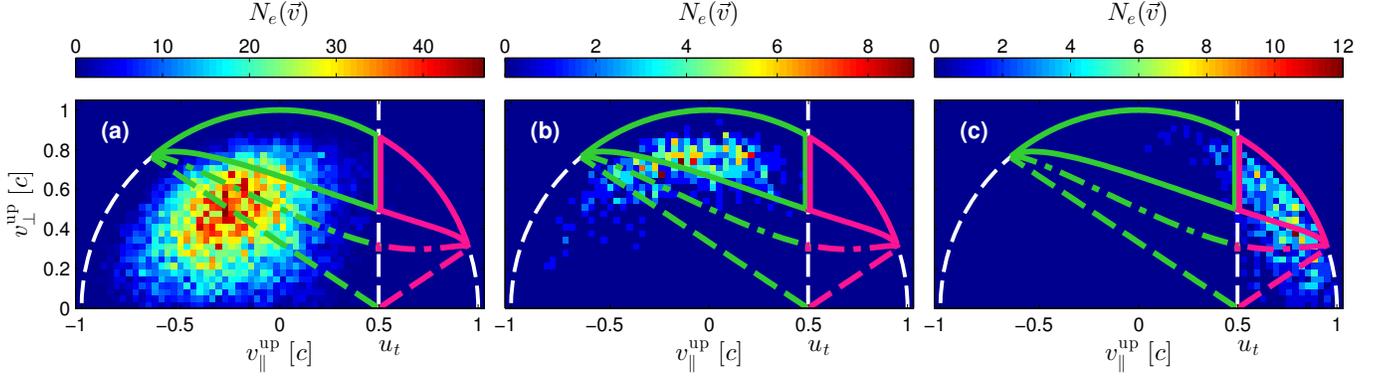}
\caption{
Velocity space $v_{\parallel}^{\rm up}-v_{\perp}^{\rm up}$ of the sample of selected electrons at different stages of their evolution. The electrons are initialized in the reference run at $\omega_{pe}t=450\ (\Omega_{ci}t = 0.83)$. 
In all the plots, the dashed half-circle indicates the speed of light. 
Panel (a) shows the velocity space of all the electrons that reach the shock, at the time when they first approach within
$50\ c/\omega_{pe}$ ahead of the shock.  
In panels (b) and (c) we plot, at two different times, the phase space of particles that are identified as eventually 
being reflected upstream from the shock. 
Panel (b) refers to the time when the identified electrons first approach within $50\ c/\omega_{pe}$ ahead of the shock, coming from the upstream side. 
Panel (c) refers to the time when they leave the shock region, i.e., at the time when they first move beyond $50\ c/\omega_{pe}$ away from the shock after the interaction. 
The 2D histograms of the velocity space plots are to be compared with the over-plotted predictions of the SDA theory (green and pink solid lines), computed assuming 
a magnetic compression ratio of $b=4$ and a cross-shock potential of $\Delta\phi = 0.5$. The effect of different values of the dimensionless cross-shock potential $\Delta\phi$ is illustrated
with the dash-dotted color lines ($\Delta\phi = 0.2$) and the dashed  color lines ($\Delta\phi = 0$). 
The vertical white dashed line tracks the location where $v_{\parallel}^{\rm up}=u_t$.  The nearly-horizontal green curve to its left is the minimum requirement on $v^{\rm up}_{i\perp}$ for SDA reflection (Equations \eqref{eq:betaperpHT} and \eqref{eq:bperptransform}).
The nearly-horizontal pink curve to the right of $v^{\rm up}_{\parallel}=u_t$ is the post-reflection mapping 
of the green curve (Equation \eqref{eq:mirrorref}). The region enclosed by the green solid lines indicates the allowed region for SDA reflection. The region enclosed by the pink lines indicates the predicted region where the electrons will lie after reflection.
}
\label{fig:theta63testprt}
\end{center}
\end{figure*}

Given that the motional electric field 
vanishes on both sides of the shock in the HT frame, we can  assume that the total electron energy and the magnetic moment $\mu$ are conserved, i.e., 
\begin{equation}\label{eq:energy}
\gamma^{\rm HT}(x) m_e c^2 - e\Phi^{\rm HT}(x) ={\rm const}~~,
\end{equation}
\begin{equation}\label{eq:mu}
\mu^{\rm HT}(x)\equiv \frac{\left[
p^{\rm HT}_{\perp}\left(x\right) 
\right]^2
}{2m_eB^{\rm HT}\left(x\right)} = {\rm const}~~,
\end{equation}
where the superscript HT refers to quantities in the HT frame and $p_\perp$ denotes the momentum perpendicular
to the magnetic field. 
Combining Equations \eqref{eq:energy} and \eqref{eq:mu}, we can solve for the velocity parallel 
to the magnetic field in the HT frame $v_{\parallel}^{\rm HT}$ as a function of the magnetic field amplification at the shock, the electric potential energy change,  the initial (denoted by subscript $i$) velocity perpendicular to the magnetic field $v_{i\perp}^{\rm HT}$ and the initial Lorentz factor  $\gamma_i^{\rm HT}$
\begin{equation}\label{eq:betapara}
v_{\parallel}^{\rm HT}\left(x\right)=c\,\sqrt{
1-\frac{1+\left(\gamma_i^{\rm HT}\right)^2\left(v_{i\perp}^{\rm HT}/c\right)^2\frac{B^{\rm HT}\left(x\right)}{B_0^{\rm HT}}}{\left[\Delta\phi\left(x\right)+\gamma_i^{\rm HT}\right]^2}
}~,
\end{equation}
where we have defined the dimensionless parameter $\Delta\phi = e\left[\Phi^{\rm HT}(x)-\Phi^{\rm HT}_0\right]/m_ec^2$ for notational convenience. 
For a particle to be reflected at the shock in the HT frame, it needs to move towards the shock in the first place,
which requires the initial velocity along the magnetic field 
line $v_{i\parallel}^{\rm HT}$ to be negative. 
Secondly, the reflection occurs when the parallel velocity in 
Equation \eqref{eq:betapara} vanishes. Combining these two conditions, we obtain the
reflection conditions in the HT frame
\begin{align} 
v_{i\parallel}^{\rm HT} & <0\ ,  \label{eq:betaparaHT}\\
v_{i\perp}^{\rm HT}  &\geq c\, \sqrt{
\frac{B_0^{\rm HT}}{B^{\rm HT}(x)}
\frac{\left[\gamma_i^{\rm HT}+\Delta\phi(x)\right]^2-1}{\left[\gamma_i^{\rm HT} \right]^2} 
}\ .~\label{eq:betaperpHT}
\end{align}
Alternatively, we can rewrite the second equation as a condition on the pitch angle in the HT frame, defined as  $\alpha^{\rm HT}\equiv\cos^{-1}\left(v_{\parallel}^{\rm HT}/v^{\rm HT}\right)$, 
\begin{equation}\label{eq:pitchangle}
\alpha_i^{\rm HT} \geq  \sin^{-1}
\sqrt{
\frac{B_0^{\rm HT}}{B^{\rm HT}(x)}
\frac{\left[\gamma_i^{\rm HT}+\Delta\phi(x)\right]^2-1}{\left[\gamma_i^{\rm HT} \right]^2-1}
}~.
\end{equation}
This states that only particles with pitch angle larger than $\alpha_i^{\rm HT}$ can be reflected back toward the upstream.
 The effect of a positive potential jump $\Delta\phi>0$ is to increase the minimum pitch angle needed for reflection, and thus reduce the number of particles satisfying the reflection condition. Physically, this stems from the fact that, if $\Delta\phi>0$, the electric force tends to attract the electrons into the shock, so it is harder for them to reflect upstream.

The velocities in the HT frame are related to those in the upstream rest frame by the relativistic velocity addition formulae
\begin{align}
v_{\parallel}^{\rm HT} &= \frac{v_{\parallel}^{\rm up}-u_t}{1-v_{\parallel}^{\rm up}u_t/c^2}\ ,\label{eq:bparatransform}\\
v_{\perp}^{\rm HT} &= \frac{v_{\perp}^{\rm up}}
{\gamma_t\left(1-v_{\parallel}^{\rm up}u_t/c^2\right)}\ ,\label{eq:bperptransform}
\end{align}
which imply that
\begin{equation}
\gamma^{\rm HT} = 
\gamma^{\rm up}\gamma_{t}\left(1-v_{\parallel}^{\rm up}u_{t}/c^2\right)~,
\label{eq:gamtransform}
\end{equation}
where $u_t$ is the relative velocity between the HT frame and the upstream rest frame defined in Equation \eqref{eq:betas} and  $\gamma_t=1/\sqrt{1-u_t^2/c^2}$ is the corresponding Lorentz factor. 
Applying the above Lorentz transformations to Equations \eqref{eq:betaparaHT}-\eqref{eq:pitchangle}, 
one can obtain the reflection conditions in the upstream rest frame. 

The condition on $v_{i\parallel}^{\rm up}$ is simply
\begin{equation}\label{eq:betaparaup}
v_{i\parallel}^{\rm up}<u_t~.
\end{equation}
The condition on $v_{i\perp}^{\rm up}$ is straightforward to derive but lengthty. However, we can gain some insight by considering the limit $\Delta\phi\to 0$. 
In this case, the particles would be reflected back upstream if
\begin{equation}\label{eq:betaperpup}
v_{i\perp}^{\rm up}\geq\gamma_t\left(
u_{t} - v_{i\parallel}^{\rm up}
\right)\tan\alpha_0\ ,
\end{equation}
where we have defined 
$
\alpha_0 = \sin^{-1}\left[B_0^{\rm HT}/B^{\rm HT}(x)\right]^{1/2}.
$ In the limit $\Delta\phi\to 0$, one can obtain a lower bound on the energy needed for reflection. Taking the equality in Equation (\ref{eq:betaperpup}), we find that the
lower boundary of the allowed region for SDA reflection (at $v_{i\parallel}^{\rm up}<u_t$) is described by
\begin{align}
\left(v_i^{\rm up}\right)^2  = &(v_{i\parallel}^{\rm up})^{2}+ \left(v_{i\perp}^{\rm up}\right)^{2} \nonumber \\  =&  
\left(1+\gamma_{t}^{2}\tan^2\alpha_{0}\right)(v_{i\parallel}^{\rm up})^{2}-2\,\gamma_{t}^{2}u_t v_{i\parallel}^{\rm up}\tan^2\alpha_{0}\nonumber\\
&+\gamma_{t}^{2}u_{t}^{2}\tan^2\alpha_{0}~.
\end{align}
The minimum of this quadratic equation with respect to $v_{i\parallel}^{\rm up}$ gives
a lower bound on the minimum velocity in the upstream rest frame
for SDA reflection
\begin{equation}\label{eq:minbeta}
v_{i,\rm min}^{\rm up} =u_t\,
\sqrt{\frac{ \tan^2{\alpha_0}}{\tan^2\alpha_0+1/\gamma_t^2}}\ ,
\end{equation}
which grows monotonically with $u_t$. It follows that, with increasing $u_t$, i.e., 
for higher shock velocity or magnetic obliquity, 
the minimum energy required to participate in the SDA process will increase. 
 Thus, the fraction of particles that can participate in SDA decreases with increasing $u_t$,  
and the reflection fraction drops further with increasing $\Delta\phi>0$, as we have discussed before.

After reflection (which we shall indicate with the subscript $r$), the parallel velocities of the particles are reversed in the HT frame:
\begin{equation}\label{eq:mirrorref}
v_{r\parallel }^{\rm HT} = -v_{i\parallel }^{\rm HT} \ ,\qquad
v_{r\perp }^{\rm HT} = v_{i\perp }^{\rm HT}\ .
\end{equation}
Transforming back to the upstream rest frame, we just need to switch the superscripts
and change the sign of $u_t$ in Equations \eqref{eq:bparatransform}-\eqref{eq:gamtransform}. 
The post-reflection  Lorentz factor and velocity are related to the pre-reflection values by
\begin{align}
\gamma_{r}^{\rm up} 
& = 
  \gamma_{i}^{\rm up}
\left[
1+\frac{2u_t\left(u_t - v_{i\parallel}^{\rm up}\right)}{c^2-u_t^2}
\right]\equiv  \gamma_{i}^{\rm up} \left(1+\Delta_{i\rightarrow r}\right)\ ,~\label{eq:deltagampercent}
\end{align}
\begin{align}
v_{r\parallel}^{\rm up} &= \gamma_t^2\frac{2\,u_t - v_{i\parallel}^{\rm up}\left(1+u_t^2/c^2\right)}{1+\Delta_{i\rightarrow r}}\ ,\label{eq:reflectbetapara}\\ 
v_{r\perp}^{\rm up} &=\frac{v_{i\perp}^{\rm up}}{1+\Delta_{i\rightarrow r}} \ .\label{eq:reflectbetaperp}
\end{align}
Since $v_{i\parallel}^{\rm up}<u_t$ is required for reflection (see Equation (\ref{eq:betaparaup})), the term
in the square brackets in Equation \eqref{eq:deltagampercent} is always larger than unity (i.e., $\Delta_{i\rightarrow r}> 0$ ). Thus, all the 
reflected particles will gain energy. 
For the same reason, all the reflected particles will suffer a reduction in  their perpendicular velocity (see Equation \eqref{eq:reflectbetaperp}). 
Given the net increase 
in energy, we conclude that, after reflection, all the particles will have a smaller perpendicular momentum and a larger parallel momentum, 
and thus move preferentially along the magnetic field.  In addition, 
for a given $v_{i\parallel}^{\rm up}$, the fractional energy gain $\Delta_{i\rightarrow r}$
increases with increasing $u_t$. 

\subsubsection{Verification of SDA in the Simulation}
To confirm that the SDA process indeed operates in our simulation, we trace the evolution of $12800$ electrons injected into the reference run at $\omega_{pe}t=450\ (\Omega_{ci}t= 0.83)$ and check whether the properties of the reflected electrons agree with the predictions of SDA. 

The predictions of SDA are indicated in the plots of  
velocity space $v_{\parallel}^{\rm up}-v_{\perp}^{\rm up}$ in Figure \ref{fig:theta63testprt}. 
The white dashed   half-circle indicates the limit $v=c$, so only the regions within the half-circle
have physical meaning.
The vertical white dashed   line marks the condition $v_{\parallel}^{\rm up}=u_t$, see Equation (\ref{eq:betaparaup}). The nearly-horizontal  green solid  curve to its left indicates the limit in Equation \eqref{eq:betaperpHT}, once transformed to the upstream rest frame using Equation \eqref{eq:bperptransform}. Only particles with  velocity to the left of the white dashed  line are approaching the shock, and only those
with velocity within the region enclosed by the green solid lines are allowed for SDA reflection. 
The pink  solid  curve to the right of the white dashed  line
is the post-reflection mapping of the green curve (following Equations (\ref{eq:deltagampercent})-(\ref{eq:reflectbetaperp})). 
Thus, the SDA theory predicts that after reflection, the velocity of the particles should occupy the region enclosed
by the pink solid  lines.

We remark that $B_0^{\rm HT}/B^{\rm HT}(x)$ and $\Delta\phi(x)$ in Equation \eqref{eq:betaperpHT}
are both functions of position, and are likely to differ for each individual particle, depending on its location at the time of reflection. But to first order, we shall assume a constant value of the magnetic compression ratio $b\equiv B^{\rm HT}(x)/B_0^{\rm HT}$ and of the cross-shock potential $\Delta\phi$ in our SDA theory. For our reference run, we use $b=4$, which is roughly the value of magnetic field compression in the HT frame at the shock overshoot. 
The dimensionless cross-shock potential is chosen to be $\Delta\phi= 0.5$, which we estimate from Lorentz transformations of the electromagnetic fields measured in the simulation frame. As we show below, a value of $\Delta\phi= 0.5$ also yields a reflection 
fraction -- defined as the fraction of particles in the initialized Maxwellian distribution 
that satisfy the SDA reflection conditions -- which matches the simulation results. 
To show the effect of different values of $\Delta\phi$, we also plot  in Figure \ref{fig:theta63testprt}  the SDA predictions corresponding to $\Delta\phi=0$ and $\Delta\phi = 0.2$ with colored dashed lines and dot-dashed lines, respectively. As we have discussed in the previous subsection, the allowed region for reflection 
 shrinks with increasing $\Delta\phi$.

While tracing the selected sample of particles, we identify those electrons that have approached the shock within $50\ c/\omega_{pe}$. After interacting with the shock, a subset of these electrons  will be reflected upstream.
Our results are plotted 
as 2D histograms in Figure \ref{fig:theta63testprt}. 
Panel (a) shows the velocity space of all the electrons that have approached the shock, at the time when they are located at $\sim 50\ c/\omega_{pe}$ ahead of the shock. We see that no electron with $v_\parallel^{\rm up}>u_t$ has reached the vicinity of the shock, just a consequence of the fact that particles initialized with $v_\parallel^{\rm up}>u_t$ would have propagated away from the shock. Despite the uncertainties in the values of the magnetic compression ratio and of the cross-shock  potential jump, Figure \ref{fig:theta63testprt}(b)  clearly demonstrates  that only electrons satisfying the SDA reflection criteria (namely, the region enclosed by the green solid lines) will  eventually be reflected back upstream. Also, the velocity distribution of the particles after reflection (Figure  \ref{fig:theta63testprt}(c)) lies well within the region enclosed by the pink solid lines, in agreement with the  SDA predictions. 

\begin{figure}[tbp]
\begin{center}
\includegraphics[width = 0.5\textwidth]{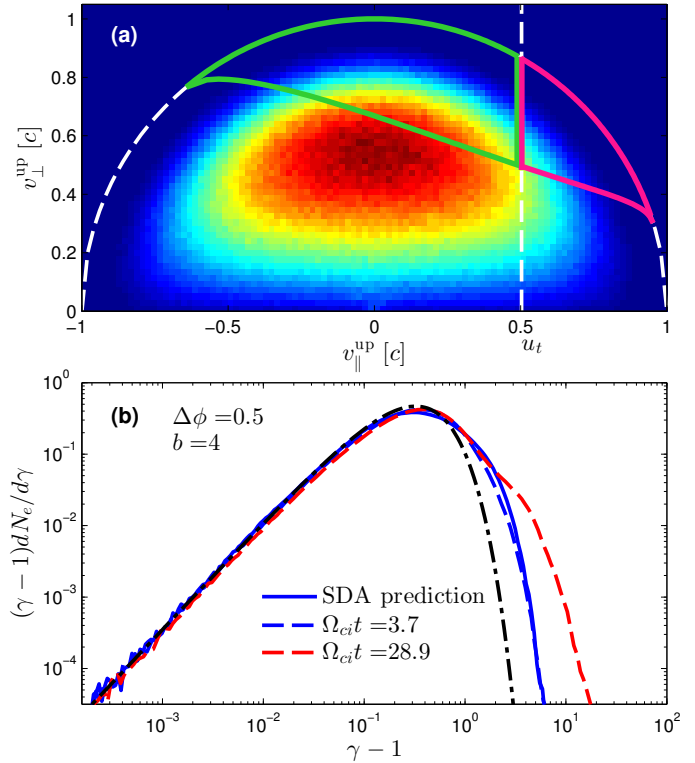}
\caption{Panel (a) shows the velocity space $v_{\parallel}^{\rm up}-v_{\perp}^{\rm up}$ of the injected electrons in the reference run. 
The SDA predictions are indicated with the green and pink solid lines as in Figure  \ref{fig:theta63testprt}, assuming $b=4$ and $\Delta\phi = 0.5$.
The synthetic spectrum is constructed by combining two populations. The first population consists of electrons with $v_{\parallel}^{\rm up}<u_t$ (i.e., in
the region to the left of the vertical white 
dashed line in panel (a)), which is required so that the electrons can interact with the shock. The second population consists of electrons that were originally within the region allowed for SDA reflection, as delimited by the green solid line. After reflection, they move to the region delimited by the pink solid line, and their energy increases according to Equation \eqref{eq:deltagampercent}. 
Panel (b): The blue solid line shows the synthetic energy spectrum computed in the way described above, following the SDA predictions.
The blue dashed line shows the electron energy spectrum measured in the
simulation at a relatively early time ($\Omega_{ci}t=3.7$)  at a distance of $50-100\ c/\omega_{pe}$ ahead of the shock. This is in very good agreement with the synthetic spectrum.
The red dashed line shows the spectrum measured at a later time ($\Omega_{ci}t = 28.9$), 
which differs significantly from the synthetic SDA spectrum. The late-time spectrum shows a 
 pronounced non-thermal component extending to much larger energies, suggesting the presence of a long-term Fermi-like acceleration mechanism.
}
\label{fig:synthetictheta63}
\end{center}
\end{figure}

As an additional test, we have used the SDA theory to compute a synthetic electron energy spectrum after one cycle of SDA,  and we have compared the synthetic spectrum to the energy spectrum measured in the simulation just upstream from the shock. 
In the upstream rest frame, the initial electron velocity distribution is a Maxwellian with 
$kT_e/m_ec^2=0.17$, which is shown in Figure \ref{fig:synthetictheta63}(a) as a 2D histogram. 
We identify the electrons that satisfy the reflection conditions (i.e., inside the  region delimited by the green lines in Figure \ref{fig:synthetictheta63}(a)), compute their 
post-reflection energy using Equation \eqref{eq:deltagampercent}, and combine them with the sub-population of the initialized electrons 
that satisfies $v_{\parallel}^{\rm up}<u_t$, so that the initialized electrons can approach the shock. Finally, we transform the energy of each electron from the upstream frame to the simulation (downstream) frame. The resulting synthetic spectrum is shown with the solid blue line in Figure  \ref{fig:synthetictheta63}(b). It agrees very well 
with the actual energy spectrum measured from the simulation at a relatively early time, $\Omega_{ci}t=3.7$ (blue dashed line). 

It should be stressed that the maximum energy of the 
predicted SDA spectrum is time-independent, for a given set of shock parameters. However, at late times, the electron energy spectrum in our simulation keeps evolving, with the non-thermal tail stretching in time to   
higher and higher energies. The red dashed line in Figure  \ref{fig:synthetictheta63}(b) shows the electron energy spectrum measured at a later time, $\Omega_{ci}t = 28.9$. The maximum energy here is $\gamma_{e,max}\gtrsim 20$, three times larger than what is predicted for one-cycle of SDA, $\gamma_{e,max}\sim 6$ (compare the red dashed line with the blue solid line). Such long-term sustained 
acceleration can only be achieved by an additional  stage of energization, akin to 
the Fermi mechanism.  

\begin{figure*}[tbp]
\begin{center}
\includegraphics[width=0.7\textwidth]{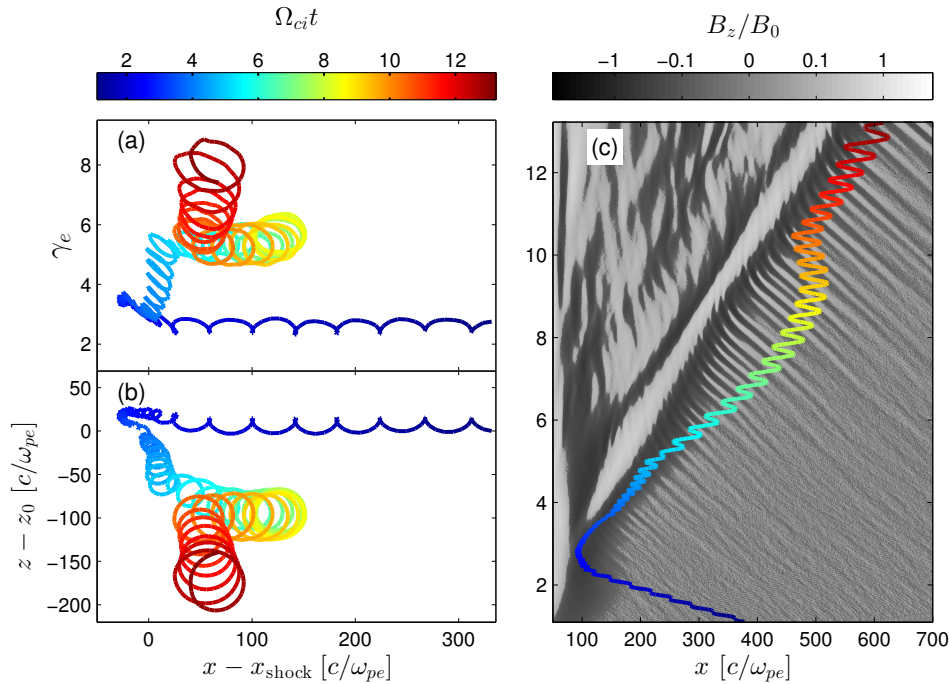}
\caption{
The evolution of a typical non-thermal electron undergoing multiple SDA cycles, in a way that resembles the standard Fermi mechanism. 
Color indicates time, from blue to red as time evolves from early to late.
Panels (a) and (b) show the evolution of the electron Lorentz factor
 and $z$-coordinate as a function 
of its distance from the shock. 
Panel (c) shows the $x$-location of the electron on top of the spatio-temporal evolution of the $y$-averaged value of $B_z/B_0$. 
All quantities are measured in the simulation (downstream) frame. 
The particle is scattered by the upstream waves plotted in panel (c) back toward the shock at $\Omega_{ci}t \sim 9$,
which allows it to gain additional energy via a second cycle of SDA. 
}
\label{fig:testprt_eg2}
\end{center}
\end{figure*}

\subsubsection{Fermi Acceleration}
\label{sec:Fermi}
In Figure \ref{fig:testprt_eg2}, we
show the evolution of a typical non-thermal electron that undergoes multiple SDA cycles, in a way resembling the well-known Fermi process. The energy and $z$-location of the particle (Figure \ref{fig:testprt_eg2}(a) and (b)) show that the electron first gains energy at  $\Omega_{ci}t\sim 2.5 - 5$ 
via SDA and is then reflected upstream. While the particle propagates upstream at $\Omega_{ci}t \sim 5-9$, it interacts
with the upstream waves (Figure  \ref{fig:testprt_eg2}(c)), generated by previous populations of returning electrons streaming ahead of the shock.
As a result of the interaction with the waves,  the electron momentum parallel to the magnetic field is reduced, 
and eventually the electron is scattered back toward the shock at $\Omega_{ci}t\sim 9$.
The interaction with the waves themselves does not yield a significant energy gain, yet
it allows the particle to approach the shock for a second time. 
When the particle reaches $50\,c/\omega_{pe}$ ahead of the shock at $\Omega_{ci}t\sim 10.5$, 
the gradient of the magnetic field confines the electron at the shock, and the particle gains energy again through  a second SDA cycle. This appears clearly from both the drift motion along $-\hat{z}$ and the corresponding increase in energy (Figure \ref{fig:testprt_eg2}(a) and (b)).

We remark that the conventional Fermi acceleration relies on scattering by waves both upstream and downstream of the shock. Although we do observe downstream waves in Figure  \ref{fig:shockflds}(a)-(c), they are not necessary for the Fermi-like acceleration we are discussing here. 
As shown in the typical particle orbit in Figure  \ref{fig:testprt_eg2}, 
the particles are reflected upstream by the magnetic field gradient 
at the shock front via the magnetic mirror effect, without penetrating the downstream region. 
The upstream waves then scatter these reflected electrons back towards the shock for additional cycles of SDA. 
We note that the particles propagate into the upstream for a distance larger than a few Larmor radii, as expected in the standard Fermi acceleration mechanism.
It is worth contrasting this behavior with that of the particle in Figure  \ref{fig:testprt_eg1}, which is not scattered back from the upstream region towards the shock. The main difference is that after the first cycle of SDA, the particle in Figure \ref{fig:testprt_eg1} returned upstream having a large component of the momentum parallel to the magnetic field. It propagated away from the shock, with little deflection by the upstream waves (which were still weak, at such early times). As a result, the interaction with the waves was not sufficient to 
scatter the electron back towards the shock for further SDA energization. 

In general, particles emerging from the SDA acceleration with a small parallel 
momentum are more favorable for being scattered back toward the shock by the upstream waves. 
After being scattered, their parallel momentum is still small, 
and since their energy has increased due to the previous SDA cycle, 
they lie in the region of velocity space favorable for additional SDA reflection and acceleration. This makes this Fermi-like acceleration process -- composed of multiple SDA cycles -- extremely efficient.
We also note that particles emerging from the SDA acceleration with a large parallel 
momentum, such as the one in Figure  \ref{fig:testprt_eg1}, though not likely to undergo Fermi acceleration, are still very important for the overall acceleration process. In fact, their large parallel momentum 
contributes to the electron temperature anisotropy in the upstream region (see Figure \ref{fig:egshock}(l)), which governs the growth of the upstream waves that scatter later generations of reflected electrons. The nature of the electron self-generated waves will be discussed in a forthcoming paper \citep{Guoinprep}. We stress that the electron self-generated waves exist and mediate efficient electron acceleration in low Mach number shocks for a wide range of physical parameters.
As we will show in a forthcoming paper \citep{Guoinprep}, we reach similar conclusions across nearly all the magnetic obliquity angles, in the temperature range $T_e = 10^7 - 10^9 K$ and for magnetizations $\sigma = 0.01-0.03$. 
The excitation of the waves requires 
the parallel electron thermal pressure, $P_{e\parallel}\equiv n_e k_B T_{e\parallel}$, to be larger than 
the magnetic pressure. In addition, it relies on the free energy provided by 
the electron temperature anisotropy $\left( T_{e\parallel}>T_{e\perp}\right)$ introduced by 
the returning electrons, which further depends on the SDA process. The detailed dependence on various physical parameters shall be presented in the forthcoming paper \citep{Guoinprep}.

\section{Summary and Discussion}\label{sec:summary}
In this work, we study from first principles the physics of electron acceleration in a low Mach number ($M_s = 3$) shock, by means of  fully kinetic PIC plasma simulations.
In our simulation, the upstream plasma follows a Maxwell-J\"uttner distribution with temperature $T_e=T_i = 10^9\ K$ and a low magnetization parameter $\sigma = 0.03$ (equivalent to a plasma beta $\beta_p = 20$). 
The upstream magnetic field is oriented at an angle of $\theta_B = 63^\circ$ with respect to the shock normal.
The physical parameters we choose are applicable to the bow shocks expected to form ahead of the G2 cloud \citep{Naryan2012,Sadowski2013} and of the S2 star \citep{Giannios2013} upon interaction with the hot accretion flow at the Galactic Center. Our parameters are also relevant to merger shocks in galaxy clusters, aside from the lower temperature of the intra-cluster plasma ($T\sim10^7\ K$). 
A complete investigation of the parameter space will be presented in a forthcoming paper \citep{Guoinprep}, where we will show explicitly that out results can be generalized to lower temperatures. 
We emphasize that, in the upstream frame, the plasma is initialized according to the physically-grounded Maxwellian distribution, instead of the so-called ``$\kappa$-distribution'' that was employed by, e.g., \cite{Park2013}.
The latter distribution contains an additional supra-thermal tail that can artificially enhance the  injection of electrons into the acceleration process.

We find that ions are not efficiently accelerated. 
In contrast, about $15\%$ of the incoming thermal electrons are accelerated up to non-thermal energies. The upstream electron energy spectrum develops a non-thermal power-law tail with slope $p\equiv -d\log N/d \log (\gamma-1)\simeq2.4$. The energy density
carried by the high-energy electrons is $\simeq 10\%$ of the bulk kinetic energy density of the incoming ions. 
The spectral cut-off energy of the upstream electron spectrum steadily 
grows with time, indicating that the acceleration process persists to late times.
The radio synchrotron spectral index expected for a slope  $p\simeq2.4$ of the electron energy distribution is 
$\alpha \equiv d\log F_\nu/d\log \nu= (1 - p)/2\simeq -0.7$ \citep{Rybicki1979}, which agrees with the radio spectral index ($\alpha = -0.6\pm 0.05$)
observed at the shock front of the radio relic in the galaxy cluster CIZA J2242.8+5301 \citep{vanWeeren2010}. 
Incidentally, the polarization analysis of the radio relic shows that the magnetic field is quasi-perpendicular, which is consistent with our setup.

We study in detail the electron acceleration mechanism. We find that
 shock drift acceleration (SDA)  governs the injection of  electrons into a Fermi-like acceleration process, that self-consistently persists in the long-term evolution of the shock. We develop a fully-relativistic theory of the SDA process and we compare it to the results of our simulation, finding excellent agreement. During the SDA process, a fraction of the incoming  electrons gain energy from 
the shock motional electric field while drifting along the shock 
surface, and they are reflected back upstream. By tracing electrons from the simulation, we demonstrate that our SDA theory properly predicts the conditions required for participating in the SDA process (and so, for the subsequent reflection toward the upstream).
We also show that the electron energy spectrum predicted by our SDA formalism, assuming one cycle of SDA acceleration, agrees well with the spectrum measured from the simulation at early times.  
However, the spectrum from the simulation at late times clearly indicates the 
existence of additional energization, beyond a single cycle of SDA. The additional energy gain
is mediated by the upstream waves self-generated by the
 electrons streaming ahead of the shock after the SDA phase. 
The upstream waves are not primarily driving the electron energy gain. Rather, they scatter the electrons propagating upstream back toward the shock for multiple cycles of SDA, thus sustaining a long-term acceleration process akin to the Fermi mechanism. 

Our study offers a possible solution to the electron 
injection problem in the low Mach number shocks present in galaxy clusters.  The bright radio luminosity that is observed from radio relics in the outskirts of galaxy clusters  seems 
to be in contradiction with the poor electron acceleration efficiency expected on theoretical grounds \citep[e.g.][for a review]{Brunetti2014}. 
Most theoretical models assume that the particles are injected via the so-called ``thermal leakage'' process 
 \citep{Malkov1998,Gieseler2000,Kang2002}, i.e., supra-thermal particles 
can propagate from the downstream back into the upstream and get injected into the Fermi process. This model requires that the electrons should have a momentum at least 
a few times larger than the characteristic post-shock  ion thermal momentum, in order to be injected into the Fermi process. 
The minimum momentum for injection increases with decreasing Mach number and is at least a factor of $\sqrt{m_i/m_e}$
larger than the expected post-shock electron momentum \citep{Kang2010}. As a result, the fraction of electrons that can participate in the Fermi process at low Mach number shocks is expected to be extremely small \citep{Kang2010}. 
\footnote{While the thermal leakage model was originally invoked for quasi-parallel shocks and has not been well developed for quasi-perpendicular shocks, it has been shown that the minimum momentum required for injection in quasi-perpendicular shocks is no less than in quasi-parallel shocks \citep[see e.g.][in the context of interplanetary shocks]{Zank2006}, so the injection efficiency in quasi-perpendicular shocks is expected to be even smaller.} In contrast, the observed bright radio emission from radio relics requires a large number of accelerated electrons.
\cite{Kang2014} propose to resolve this conflict by assuming the existence of electrons following a $\kappa$-distribution, which has an {\it ad hoc} supra-thermal tail with a power-law shape. However, the existence of such supra-thermal electrons at the outskirts of galaxy clusters has never been demonstrated \citep{Pinzke2013}.

\begin{figure*}[tbp]
\begin{center}
\includegraphics[width = 0.95\textwidth]{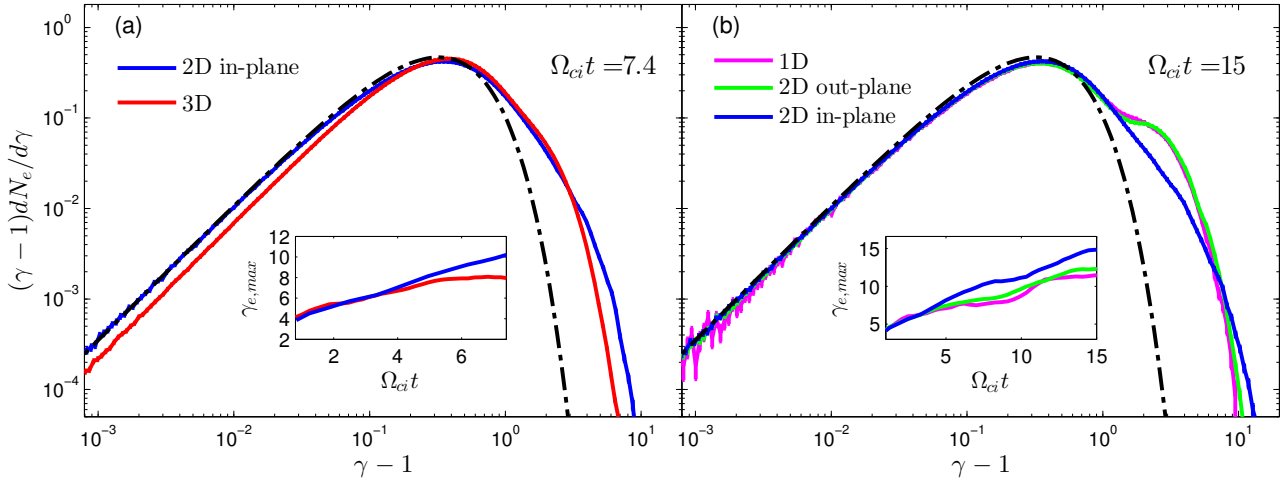}
\caption{
Panel (a) compares the electron energy spectra from the reference run (2D in-plane) and 
the 3D run measured at $\Omega_{ci}t = 7.4$ at a distance of $60-160\ c/\omega_{pe}$ ahead of the shock.
Panel (b) compares the electron energy spectra from the 1D run, the reference run (2D in-plane)
and the 2D out-plane run measured at $\Omega_{ci}t = 15$ at a distance of $60-160\ c/\omega_{pe}$ ahead of the shock.
}
\label{fig:spec_1d2d3d}
\end{center}
\end{figure*}

The key issue of the thermal-leakage model summarized above is that 
it assumes that the electrons, in order to be injected into the Fermi process, need to 
be scattered by the MHD waves in the downstream region to propagate back into the upstream, and thus they 
have to possess very large momenta in the first place (so that their Larmor radius is larger than the scale of the MHD turbulence). 
In contrast, our mechanism, based on first-principle PIC simulations, does not involve 
any scattering in the downstream turbulence. Rather,  
the shock itself acts as a magnetic mirror, reflecting a fraction of the incoming electrons back upstream after the SDA stage. 
The minimum electron momentum required for reflection via the SDA mechanism is much lower (by a factor of $\sim m_e/m_i$)
than that required in the thermal-leakage model.
We emphasize that, while the electron injection efficiency in the thermal-leakage model should significantly decrease for higher mass ratios, our results are insensitive to the choice of $m_i/m_e$, as shown in Appendix \ref{sec:massratio}.
In addition, in our study we show that the electrons propagating upstream can self-generate magnetic waves, without any need for the MHD turbulence that is usually invoked in the thermal-leakage models described above.
 
We also remark that the SDA-mediated injection that we describe in this work is complementary to the injection mechanisms invoked in high Mach number and low plasma beta shocks, which have been extensively studied in application to Supernova Remnants.
The often invoked electron shock surfing acceleration cannot operate in low Mach number 
shocks, since the  Buneman instability is suppressed in hot plasmas. On the other hand, SDA cannot operate efficiently in high Mach number shocks, because the fraction of velocity space that allows injection via SDA is shrinking with increasing Mach number, resulting in poor acceleration efficiencies. 
The injection by whistler waves proposed by \cite{Riquelme2011}
in low beta flows is not playing any important role in high plasma beta shocks, where energization via SDA is dominating.
On the other hand, the magnetic waves that allow for multiple SDA cycles in our mechanism are suppressed in low plasma beta flows, as will be demonstrated in a forthcoming paper \citep{Guoinprep}.

\acknowledgements
\section*{Acknowledgements}
X.G. and R.N. are supported in part by NASA grant NNX14AB47G.
X.G. thanks Philip Mocz for helpful comments on the manuscript and Pierre Christian for useful discussions. We thank R. van Weeren, A. Spitkovsky, and J. Park for helpful comments.
L.S. is supported by NASA through Einstein
Postdoctoral Fellowship grant number PF1-120090 awarded by the Chandra
X-ray Center, which is operated by the Smithsonian Astrophysical
Observatory for NASA under contract NAS8-03060. 
The computations in this paper were run on the Odyssey cluster supported by the FAS Division of Science, Research Computing Group at Harvard University.

\appendix
\section{Dependence on the dimensionality}\label{sec:2d3d}
Our reference run is performed on a 2D spatial domain, although we solve for all the 3D components of the 
particle momenta and of the electromagnetic fields. To test the validity of our choice of a reduced dimensionality, we have run 
simulations with the same physical parameters as in our reference run, but in 1D and 3D computational domains. Similarly, to validate our choice for   the orientation of the magnetic field (lying in the simulation plane, i.e., $\varphi_B = 0^\circ$  in our reference run),  we have performed a 2D simulation with $\varphi_B = 90^\circ$ (referred to as 2D out-plane). 
For the 1D run, the box size along the $y$-dimension is only $1\, c/\omega_{pe}$. For the 3D run, the box size 
along the $z$-dimension is equal to that along the $y$-dimension, which is $76\ c/\omega_{pe}$ (corresponding to $7.6\ c/\omega_{pi}$, or almost one ion Larmor radius).
In our 3D run, we choose to resolve
the electron skin depth with 5 cells, and we initialize the upstream plasma with 2 computational particles per cell (one per species), as opposed to the larger value (32) that we can employ in 2D simulations.

In Figure \ref{fig:spec_1d2d3d}(a), we compare the electron energy spectra from 
the reference run (2D in-plane) and the 3D run at $\Omega_{ci}t = 7.4$, 
the latest stage of evolution of the 3D simulation. We find that the energy spectra agree very well in terms of both the normalization and the power-law slope of the non-thermal tail. However the spectrum cuts off at a lower energy in the 3D run, 
and the maximum energy also grows slower at late times. This is likely an artifact of the 
lower number of computational particles per cell (2 in the 3D run versus 32 in the reference run). We have checked that the upstream waves in 3D show a similar pattern as in our reference run. 
Thus, Fermi acceleration is expected to operate in the same way in 2D and 3D. 

The comparison of the electron energy spectra between the reference run, the 1D run 
and the 2D out-plane run (Figure \ref{fig:spec_1d2d3d}(b)) shows that the spectra from 
both the 2D out-plane run and the 1D simulation present an artificially higher normalization than those from the 3D and 2D in-plane runs. 
The growth of the maximum energy is also slower than in the reference run, as shown in the subpanel. 
This indicates that the 2D out-plane configuration misses some of the important physics. 
Our conclusion is that the 2D in-plane configuration is a good choice to capture the acceleration physics
of the full 3D problem.

\section{Dependence on the mass ratio}\label{sec:massratio}
\begin{figure}[tbp]
\begin{center}
\includegraphics[width = 0.5\textwidth]{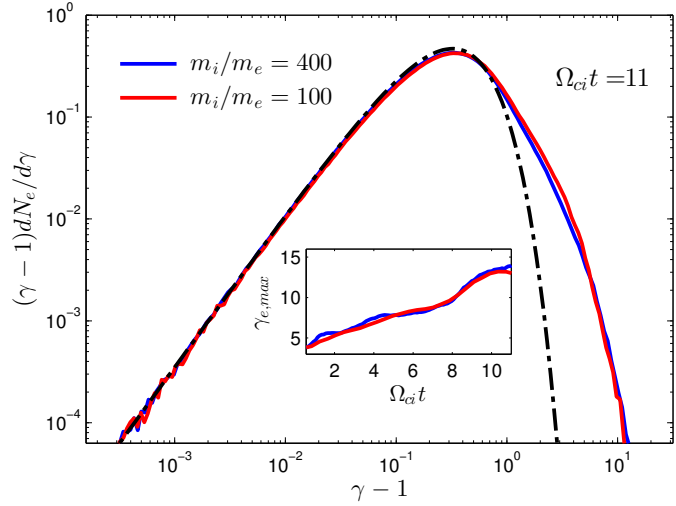}
\caption{Electron energy spectra and the evolution of the maximum Lorentz factor from the run with $m_i/m_e = 400$ (blue) and the reference run (red) measured at $\Omega_{ci}t = 11$ at a distance of $60-160\ c/\omega_{pe}$
ahead of the shock.
The spectra agree very well, as expected when we change the value of the mass ratio but keeping $u_t$ fixed (Equation \eqref{eq:betas2}). }
\label{fig:massratio}
\end{center}
\end{figure}

Since our simulations employ a reduced mass-ratio $m_i/m_e = 100$, it is natural to wonder
how the results will change when a realistic mass-ratio of $m_i/m_e = 1836$ is used. 
We note that $u_t$ and the dimensionless electric potential jump $\Delta \phi$ are the key parameters that determine the minimum energy required 
for participating in the SDA process (Equation \eqref{eq:minbeta} and discussion thereafter). Their value controls
the fraction of reflected electrons and the maximum energy gain  per cycle (Equation \eqref{eq:deltagampercent}). 
To understand the effect of the mass ratio, we rewrite $u_t$ and $\Delta \phi\sim m_i u_0^2/(2 m_e c^2)$ in terms of the simulation parameters as
\begin{eqnarray}\label{eq:betas2}
u_t & = & u_{\rm sh}^{\rm up}\sec\theta_B =\sqrt{\frac{2\Gamma k_B T_i}{m_i}} M_s \sec\theta_B=\sqrt{\frac{2\Gamma k_BT_e}{m_e } }\sqrt{\frac{m_e}{m_i}}M_s\sec\theta_B \ ,\\
\Delta \phi & \sim &  \frac{m_i u_0^2}{2\,m_e c^2}=M^2 \Gamma \frac{k_B T_e}{m_e c^2}\ .
\end{eqnarray}
This suggests that, for fixed $T_e=T_i$, $M$ and $M_s$,  if we scale 
the obliquity angle $\theta_B$ such that $\sec\theta_B\sqrt{m_e/m_i}=\rm const$, the acceleration efficiency should be unchanged.  
Since the relevant time scale for SDA and the shock evolution is $\Omega_{ci}^{-1}$, we should expect the spectra to be comparable at the same time in units of $\Omega_{ci}^{-1}$.

To test this prediction, we have run a simulation with the same 
physical parameters as in our reference run but with a larger mass ratio, $m_i/m_e=400$.  Correspondingly, we have changed the field obliquity to $\theta_B=77^\circ$, which ensures that the value of $u_t$ is unchanged. 
Figure \ref{fig:massratio} shows the comparison at $\Omega_{ci}t = 11$ of the upstream electron energy spectra, between $m_i/m_e=100$ (red) and $m_i/m_e=400$ (blue). We see excellent agreement, as expected. This confirms that the efficiency of both the SDA mechanism and of the Fermi acceleration process is independent of the mass ratio, once the magnetic obliquity is properly rescaled. In addition, we observe that the upstream waves for $m_i/m_e=400$ show a similar pattern as in our reference run.

We then conclude that our results, that employ a reduced mass ratio, can be generalized to the realistic mass ratio. For $m_i/m_e = 1836$, the 
corresponding field obliquity is $\theta_B=84^\circ$. 
For this value of $\theta_B$, 
electron acceleration should proceed exactly in the same way as we have presented here for our reference run with $m_i/m_e = 100$. 
The independence of our mechanism on the mass ratio is to be contrasted with the conclusions by \cite{Riquelme2011}, finding 
that electron injection by oblique whistler waves in \textit{low plasma beta} shocks depends sensitively on the mass ratio. In our \textit{high plasma beta} shocks, the strength of the magnetic waves mediating electron acceleration does not depend on mass ratio, as we will show in the forthcoming paper \citep{Guoinprep}. In addition, our mechanism does not rely on direct acceleration by the electric field of the waves, as opposed to the work by \cite{Riquelme2011}.

\bibliographystyle{apj}
\bibliography{accel}

\end{document}